\documentclass[aip, floatfix, jcp, reprint, longbibliography]{revtex4-1}

\usepackage{enumitem}
\usepackage{graphicx}          
\usepackage{amsmath}    
\usepackage{amssymb}    
\usepackage{amsthm}
\usepackage[]{algorithm}
\usepackage{algpseudocode}
\usepackage{enumitem}
\usepackage{graphicx}   
\usepackage{subfigure}  
\usepackage{hyperref}   
\usepackage[capitalise]{cleveref}   
\usepackage{appendix}
\usepackage{bm}
\usepackage{color, soul}

\begin{document}

\title{Deflation reveals dynamical structure in nondominant reaction coordinates}

\author{Brooke E. Husic}
\email{bhusic@stanford.edu}
\affiliation{Department of Mathematics and Computer Science, Freie Universit{\"a}t, Berlin, Germany}
\affiliation{Department of Chemistry, Stanford University, Stanford, CA, 94305, USA}
\author{Frank No{\'e}}
\email{frank.noe@fu-berlin.de}
\affiliation{Department of Mathematics and Computer Science, Freie Universit{\"a}t, Berlin, Germany}
\affiliation{Department of Chemistry, Rice University, Houston, TX, 77005, USA}


\begin{abstract}
The output of molecular dynamics simulations is high-dimensional, and the degrees of freedom among the atoms are related in intricate ways.
Therefore, a variety of analysis frameworks have been introduced in order to distill complex motions into lower-dimensional representations that model the system dynamics.
These dynamical models have been developed to optimally approximate the system's global kinetics.
However, the separate aims of optimizing global kinetics and modeling a process of interest diverge when the process of interest is not the slowest process in the system.
Here, we introduce deflation into state-of-the-art methods in molecular kinetics in order to preserve the use of variational optimization tools when the slowest dynamical mode is not the same as the one we seek to model and understand.
First, we showcase deflation for a simple toy system and introduce the deflated variational approach to Markov processes (dVAMP).
Using dVAMP, we show that nondominant reaction coordinates produced using deflation are more informative than their counterparts generated without deflation.
Then, we examine a protein folding system in which the slowest dynamical mode is not folding.
Following a dVAMP analysis, we show that deflation can be used to obscure this undesired slow process from a kinetic model, in this case a VAMPnet.
The incorporation of deflation into current methods opens the door for enhanced sampling strategies and more flexible, targeted model building.
\end{abstract}
\maketitle


\section{Introduction}

Molecular dynamics (MD) simulations provide a detailed view of a system that describes its atomistic dynamics:~how a protein folds or misfolds, how a ligand binds, or how conformational changes occur.
However, interpretable dynamical motifs are encoded in collective variables that must be computed from the simulation output.
A wealth of research has been dedicated to extracting thermodynamic and kinetic information from time-series simulation data, often under the assumption that a Markovian lag time can be defined that is long enough for transitions to be memoryless but short enough to resolve the dynamics of interest (see Ref.~\onlinecite{husic2018markov} for a review).

When models are Markovian, the future state of the system depends only on its present state, and not on its history. Thus, the system dynamics can be completely described by \emph{time-lagged pairs} of data points.
The true Markovian dynamics are described by some unknown dynamical operator whose eigen- or singular functions can be approximated in a data-driven way using the time-lagged representation~\cite{schutte2015critical}.

Recently, two variational principles were developed for computing the \emph{optimal} approximations to the dominant eigen- or singular functions of the unknown operator, which represent the slow dynamical modes of the system.
First,~\citet{noe2013variational} introduced the variational approach to conformational dynamics (VAC) for reversible dynamics; more recently,~\citet{wu2017vamp} derived the variational approach to Markov processes (VAMP) for irreversible dynamics, which is a generalization of the VAC.

When analyzing time-lagged data, suitable \emph{feature} representations must be chosen for the data, which often involve transformations of the original spatial coordinates such that the features are rotation- and translation-invariant.
For proteins, common features include backbone dihedral angles and pairwise contact distances.
From the time-lagged feature representation, VAMP can be used in two ways.
First, VAMP yields the optimal \emph{linear} model from those features.
Second, VAMP can be used to compare the approximation quality of nonlinear models of the features, such as Markov state models (MSMs)~\cite{husic2018markov}, Koopman models~\cite{wu2017variational}, or VAMPnets~\cite{mardt2018vampnets}.
A method for comparing feature choices directly has also recently been presented~\cite{scherer2019variational}.

The introduction of variational methods into kinetic model construction has shifted the modeling community away from heuristic parameter choices and towards objective model rankings.
However, the VAC and the VAMP optimize the global kinetics, which are characterized by the slowest dynamical modes of the system.
It is a common criticism of these methods that the slow modes are not always the same as the modes of interest~\cite{husic2018markov, scherer2019variational}.
For example, the optimization method introduced by~\citet{banushkina2015nonparametric} identifies a rare event as the optimal reaction coordinate instead of the desired mode, and the method derived by~\citet{mcgibbon2017identification} applied to a protein folding dataset finds an irrelevant distance contraction instead of the folding process.

Thus, applying a variational method when modeling a time-series dataset will optimize the approximation of the slowest modes, regardless of whether they are interesting.
While in some cases it can be straightforward to modify the kinetic model \emph{a posteriori} in order to remove the undesired modes after model construction, it becomes messier if not impossible to remove the same slow mode from every candidate model when multiple models are ranked and their modes are nonequivalent.

Furthermore, constraints of commonly used kinetic models require that the slow modes they identify are mutually orthogonal, which places restrictions on the modeling of the process of interest if it is not the slowest process.
Finally, researchers sometimes find that the reaction coordinates of desired modes feature structures also present in slower reaction coordinates of undesired modes, thus obscuring the nondominant process.

Inspired by these limitations, in this work we consider the use of \emph{deflation} to address two problems at the frontier of molecular kinetics analyses.
First, we investigate whether the incorporation of deflation into the variational approach can increase the interpretability of nondominant reaction coordinates.
Deflated reaction coordinates are no longer required to be orthogonal to previously computed coordinates; thus, if previously computed reaction coordinates are undesired, deflation may prevent the contamination of interesting modes with uninteresting ones.
Second, we examine whether the deflation of a specific process can be used construct a new kinetic model that optimizes the description of only the desired slow dynamics.
In this case, deflation is used to create a new basis, from which multiple models can be constructed and ranked with the guarantee that they are not optimizing for the accuracy of the uninteresting slow mode.
From our results, we suggest several next steps for sampling and model construction.


\section{Theory} \label{sec:theory}

\subsection{Time-lagged independent component analysis} \label{sec:tica}

The bio- and chemical physics communities often analyze time-series data in which the ordering of data points is not arbitrary but instead is indexed by time.
A suite of algorithms developed for MD simulations involves the analysis of time-lagged data using feature transformations such that the dynamics are Markovian---i.e., independent of the system history for a defined lag time $\tau$.
Since the dynamics do not depend on system history, the entire dataset can be represented by $T - \tau$ time-lagged pairs, for $T$ total simulation time.

Equivalently, time-lagged data can be represented by two matrices,

\begin{align}
    \mathbf{X} &\equiv [\mathbf{z}_1, \dots, \mathbf{z}_{T-\tau}]^\top  \label{eq:x}, \\
    \mathbf{Y} &\equiv [\mathbf{z}_\tau, \dots, \mathbf{z}_{T}]^\top \label{eq:y},
\end{align}

\noindent{}where $\mathbf{z}_t$ is the column vector of mean-free feature values at time $t$, where $\mathbf{X}$ represents the first $T-\tau$ data points and $\mathbf{Y}$ represents their values after a time shift of $\tau$.
For MD analyses, we assume that the length of $\mathbf{z}_t$ is smaller than $T$; i.e., $\mathbf{X}$ and $\mathbf{Y}$ have more rows than columns.
For example, $\mathbf{z}_t$ might contain the (mean-free) inter-residue distances or side chain torsional angles of a protein.

Time-lagged independent component analysis (TICA)\cite{naritomi2011slow,schwantes2013improvements,perez2013identification} leverages the matrix representations~\eqref{eq:x} and~\eqref{eq:y} to solve the generalized eigenvalue problem,

\begin{align}
    \mathbf{C}_{\tau} \mathbf{\Phi} = \mathbf{C}_{0} \mathbf{\Phi} \mathbf{\Lambda}, \label{eq:tica_gev}
\end{align}

\noindent{}for symmetric correlation and covariance matrices $\mathbf{C}_\tau$ and $\mathbf{C}_0$, where,

\begin{align}
\mathbf{C}_\tau &\equiv \frac{\mathbf{X}^\top\mathbf{Y} + \mathbf{Y}^\top\mathbf{X}}{2(T-\tau)}, \label{eq:tica_corr}\\
\mathbf{C}_0 &\equiv \frac{\mathbf{X}^\top\mathbf{X} + \mathbf{Y}^\top\mathbf{Y}}{2(T-\tau)}, \label{eq:tica_cov}
\end{align}

\noindent{}respectively.

The TICA solution~\eqref{eq:tica_gev} gives a set of \emph{weights} stored in the columns of $\mathbf{\Phi}$ (i.e., the eigenvector solutions) which can be used to transform the data from its original space to a new space.
Each solution is a \emph{reaction coordinate}\cite{mcgibbon2017identification}:~it characterizes a dynamical process within the data.
We will see later that TICA is a special case of canonical correlation analysis (CCA).

It is clear from definitions~\eqref{eq:tica_corr} and~\eqref{eq:tica_cov} that a strict reversibility requirement is imposed upon the TICA formulation.
In particular,~\eqref{eq:tica_corr} assumes that viewing the data in the backwards direction in time would be equally valid if it were to have occurred in the forward direction, which introduces bias that may not be appropriate for systems with rare events.
This reversibility requirement has both mathematical and physical implications:~the eigenvalues of the solution will be real and the dynamical system modeled is assumed to be in thermodynamic equilibrium.
The two interpretations are connected:~the processes in the simulated system, represented by the weights (eigenvector solutions), relax to equilibrium when the dynamics are reversible.

TICA was introduced into MD analysis in 2011 as a method to identify slow kinetic modes in  proteins~\cite{naritomi2011slow}.
Later, TICA was used as a preprocessing step in the construction of MSMs~\cite{schwantes2013improvements}.
Most importantly for our purposes, TICA is equivalent to the linear VAC:~namely, it has been shown to produce the \emph{optimal linear approximations} for the slow modes of reversible systems~\cite{perez2013identification}.
In other words, by transforming our time-lagged features using TICA solutions, we obtain the best possible linear description of the system kinetics using those features---as long as we are comfortable assuming reversible dynamics.

\subsection{Variational approach to Markov processes}

However, what if we are not comfortable assuming reversible dynamics, but still want to obtain linearly optimal approximations of the slow modes in our features?
In 2017, \citet{wu2017vamp} solved this problem:~instead of seeking the eigenvector solutions of TICA, which requires a reversibility assumption, the optimal solutions instead come from the \emph{singular} vectors of the so-called Koopman matrix.

The Koopman matrix $\mathbf{K}$ minimizes the error in the regression problem $\mathbf{Y} = \mathbf{XK}$.
For time-lagged data, $\mathbf{K}$ thus approximates a dynamical propagator:~the right hand side approximates the propagation of the data after the lag time $\tau$.\footnote{Here, we adopt the statistics convention by writing our linear equation $\mathbf{Y} = \mathbf{X}\mathbf{K}$. In the propagator context, it is more common to write $\mathbf{Y}=\mathbf{K}\mathbf{X}$, and thus the equations for $\mathbf{K}$ in this work are equal to $\mathbf{K}^\top$ with the latter convention.}
Approximating the singular functions of the propagator by analyzing $\mathbf{K}$ is the linear VAMP~\cite{wu2017vamp, paul2019identification}.

To compute $\mathbf{K}$ we first calculate three matrices from our mean-free, time-lagged dataset (without any reversibility requirements):

\begin{align}
    \mathbf{C}_{00} &\equiv \frac{1}{T-\tau} \mathbf{X}^\top\mathbf{X}, \label{eq:c00}\\
    \mathbf{C}_{0\tau} &\equiv \frac{1}{T-\tau} \mathbf{X}^\top\mathbf{Y},  \label{eq:c0t}\\
    \mathbf{C}_{\tau\tau} &\equiv \frac{1}{T-\tau} \mathbf{Y}^\top\mathbf{Y}. \label{eq:ctt}
\end{align}

\noindent{}Then, we perform a whitening transformation on the matrices $\mathbf{X}$ and $\mathbf{Y}$,

\begin{align}
    \tilde{\mathbf{X}} &\equiv \mathbf{XC}_{00}^{-\frac{1}{2}}, \label{eq:xtilde}\\
    \tilde{\mathbf{Y}} &\equiv \mathbf{YC}_{\tau\tau}^{-\frac{1}{2}}, \label{eq:ytilde}
\end{align}

\noindent{}in order to remove their internal covariances. 
Instead of determining $\mathbf{K}$ directly, we instead calculate $\tilde{\mathbf{K}}$, which propagates $\tilde{\mathbf{X}} \rightarrow \tilde{\mathbf{Y}}$ in the \emph{whitened} space.

The solution to the regression problem $\tilde{\mathbf{Y}} = \tilde{\mathbf{X}}\tilde{\mathbf{K}}$ is given by the ordinary least squares estimator\cite{wehmeyer2018time},

\begin{align}
\tilde{\mathbf{K}} = (\tilde{\mathbf{X}}^\top\tilde{\mathbf{X}})^{-1}\tilde{\mathbf{X}}^\top\tilde{\mathbf{Y}},
\end{align}

\noindent{}which, by using substitutions from~\eqref{eq:c00} through \eqref{eq:ytilde}, we can show yields\footnote{We assume in~
\eqref{eq:xtilde}-\eqref{eq:ktilde} that the matrices $\mathbf{X}^\top\mathbf{X}$, etc., are full rank.},

\begin{align}
    \tilde{\mathbf{K}} &= (\mathbf{X}^\top\mathbf{X})^{-\frac{1}{2}} \mathbf{X}^\top\mathbf{Y} (\mathbf{Y}^\top\mathbf{Y})^{-\frac{1}{2}} =  \mathbf{C}_{00}^{-\frac{1}{2}}\mathbf{C}_{0\tau}\mathbf{C}_{\tau\tau}^{-\frac{1}{2}}. \label{eq:ktilde}
\end{align}

VAMP proceeds by performing the singular value decomposition (SVD) $\tilde{\mathbf{K}} = \mathbf{U'SV'}^\top$, and then transforming the resulting matrices of left ($\mathbf{U'}$)  and right ($\mathbf{V'}$) singular values into the non-whitened space.
This procedure is summarized in Alg.~\ref{alg:vamp}.

\begin{algorithm}[H]
\begin{algorithmic}[1]
\Require $\mathbf{X} \in \mathbb{R}^{n \times m}, \mathbf{Y} \in \mathbb{R}^{n \times p}$
\Comment{mean-free columns}
\State $\mathbf{U'SV'}^\top \gets (\mathbf{X}^\top\mathbf{X})^{-\frac{1}{2}} \mathbf{X}^\top\mathbf{Y} (\mathbf{Y}^\top\mathbf{Y})^{-\frac{1}{2}}$
\State $\mathbf{U} \gets n^{\frac{1}{2}}(\mathbf{X}^\top\mathbf{X})^{-\frac{1}{2}} \mathbf{U}' $
\State $\mathbf{V} \gets n^{\frac{1}{2}} (\mathbf{Y}^\top\mathbf{Y})^{-\frac{1}{2}} \mathbf{V}' $
\State \textit{x weights} are the columns of $\mathbf{U}$
\State \textit{y weights} are the columns of $\mathbf{V}$
\caption{\label{alg:vamp}VAMP}
\end{algorithmic}
\end{algorithm}

\noindent{}The outputs $\mathbf{U}$ and $\mathbf{V}$ store the \emph{reaction coordinates} in their columns---the weights for the transformations of  $\tilde{\mathbf{X}}$ and $\tilde{\mathbf{Y}}$, respectively.
By the definition of the SVD, we have the following orthogonality conditions:

\begin{align}
    \mathbf{U}^\top\mathbf{C}_{00}\mathbf{U} &= \mathbf{I}, \label{eq:uortho}\\
    \mathbf{V}^\top\mathbf{C}_{\tau\tau}\mathbf{V} &= \mathbf{I}. \label{eq:vortho}
\end{align}

\noindent{}When $\mathbf{C}_{00} = \mathbf{C}_{\tau\tau}$ and $\mathbf{C}_{0\tau} = \mathbf{C}_{0\tau}^\top$, the SVD of $\tilde{\mathbf{K}}$ is equal to is eigendecomposition.
In this case, the data is reversible, and VAMP is identical to TICA~\cite{naritomi2011slow,schwantes2013improvements,perez2013identification} and the linear VAC~\cite{noe2013variational}.

\subsection{Iterative CCA with NIPALS}

CCA is a well-known statistical algorithm~\cite{hotelling1936relations, wold1985partial} that finds transformations of $\mathbf{X}$ and $\mathbf{Y}$ such that their representations in the new latent space (i.e., defined by the transformations) are maximally correlated.
Thus, VAMP is a version of CCA where the data is time-lagged~\cite{wu2017vamp, noe2018machine}.
For VAMP, we saw in Alg.~\ref{alg:vamp} that a single SVD is performed to obtain all the weights simultaneously.

CCA can be performed with a single SVD as in VAMP, or alternatively by using iterative deflation.
This produces \emph{different} weights due to the deflation process.
The iterative deflation algorithm developed by Wold~\cite{wold1985partial} proceeds as follows:

\begin{algorithm}[H]
\begin{algorithmic}[1]
\Require $\mathbf{X} \in \mathbb{R}^{n \times m}, \mathbf{Y} \in \mathbb{R}^{n \times p}$\Comment{mean-free columns}
\Ensure $\mathbf{A}^{+}$ indicate the Moore-Penrose pseudoinverse\footnote{The Moore-Penrose pseudoinverse $\mathbf{A}^+$ is defined as $(\mathbf{A}^\top\mathbf{A})^{-1}\mathbf{A}^\top$ and is equivalent to $\mathbf{A}^{-1}$ when $\mathbf{A}$ is invertible~\cite{penrose1955generalized}.} of $\mathbf{A}$
\For{each component $c$}
\State choose nonzero vector $\mathbf{\omega}_c$
\While{$\mathbf{\phi}_c$ is not converged}
\State $\mathbf{\phi}_c \gets \mathbf{X}^{+}\mathbf{\omega}_c$ \Comment{$\mathbf{\phi}_c \equiv c$th \textit{x weight} vector}
\State $\mathbf{\xi}_c \gets \mathbf{X}\mathbf{\phi}_c$  \Comment{$\mathbf{\xi}_c \equiv c$th \textit{x score} vector} \\
\State $\mathbf{\psi}_c \gets \mathbf{Y}^{+}\mathbf{\xi}_c$ \Comment{$\mathbf{\psi}_c \equiv c$th \textit{y weight} vector}
\State $\mathbf{\omega}_c \gets \mathbf{Y}\mathbf{\psi}_c$ \Comment{$\mathbf{\omega}_c \equiv c$th \textit{y score} vector}
\EndWhile \\
\State $\mathbf{X} \gets \mathbf{X} - \xi_c[\mathbf{X}^\top\mathbf{\xi}_c(\mathbf{\xi}_c^\top\mathbf{\xi}_c)^{-1}]^\top$ \Comment{Deflation of $\mathbf{X}$}
\State $\mathbf{Y} \gets \mathbf{Y} - \omega_c[\mathbf{Y}^\top\mathbf{\omega}_c(\mathbf{\omega}_c^\top\mathbf{\omega}_c)^{-1}]^\top$ \Comment{Deflation of $\mathbf{Y}$}
\EndFor
\caption{\label{alg:ccanipals} CCA (NIPALS with iterative deflation)/dVAMP}
\end{algorithmic}
\end{algorithm}

The nonlinear iterative least squares (NIPALS) method in lines 1--9 of Alg.~\ref{alg:ccanipals} is equivalent to the power method for computing dominant singular vectors~\cite{wold1973nonlinear,wegelin2000survey}.
In lines 11--12, the data is deflated such that new dominant singular vectors can be calculated in the next iteration.
In the molecular kinetics context, we can call Alg.~\ref{alg:ccanipals} ``dVAMP'', i.e.,~VAMP with deflation.
The first reaction coordinates (weights) calculated from dVAMP are identical to the first VAMP coordinates (see Appendices~\ref{app:intuition} and~\ref{app:ccasvd}).
The crucial difference between VAMP and dVAMP is that the \emph{nondominant} reaction coordinates (i.e., all but the first coordinate) are different due to deflation.

Deflation is common to statistical algorithms such as partial least squares (PLS) and principal component analysis (PCA)\cite{wold1975path, wegelin2000survey}, both of which can be performed using Alg.~\ref{alg:ccanipals} with changes only to lines 4 and 7.\footnote{PCA is equivalent to PLS when $\mathbf{X} = \mathbf{Y}$.}
Deflation can be viewed as subtracting the rank-one approximations of $\mathbf{X}$ and $\mathbf{Y}$ at step $c$ from $\mathbf{X}$ and $\mathbf{Y}$ at the previous update~\cite{wegelin2000survey}.
This is intended to \emph{minimize the effect} of a previously computed weight on the subsequent weights by removing the presence of the associated singular value and vector\cite{white1958computation}.
Using deflation to obtain the weights destroys the orthogonality conditions in~\eqref{eq:uortho} and~\eqref{eq:vortho}.
Furthermore, there is no persistent matrix of which the singular vectors are computed (see also Appendix~\ref{app:operator}).

By applying the deflated weights to the \emph{non-deflated data},
\begin{align}
    \xi'_c &= \mathbf{X}\phi_c, \label{eq:scorex}\\
    \omega'_c &= \mathbf{Y}\psi_c, \label{eq:scorey}
\end{align}

\noindent{}dVAMP produces transformations that differ from VAMP.\footnote{
Note that if the deflated weights are applied to the the deflated data, the transformations will be the same in VAMP and dVAMP. The nonequivalence of VAMP and dVAMP transformations occurs when applying the deflated weights to \emph{nondeflated} data.
}
We expect the weights identified using Alg.~\ref{alg:ccanipals}
to be minimally contaminated by previously identified coordinates.

To demonstrate the application of dVAMP in elucidating nondominant reaction coordinates, we consider a low-dimensional toy system.
Then, for an atomistic protein folding system, we explore the use of deflation by performing a dVAMP analysis in order to deflate a \emph{specific} slow dynamical mode from the feature data \emph{before} applying a kinetic model.
This strategy allows selected modes to be obscured from the model optimization procedure, because the associated timescale no longer appears slow.


\section{Experiments} \label{sec:experiments}

\subsection{Asymmetric double wells} \label{subsec:dw}

Due to deflation, we expect that reaction coordinates obtained with dVAMP will minimally contaminate subsequently calculated coordinates.
To investigate this, we design a dataset with dependencies among its degrees of freedom in order to mimic a real system, but with few enough dimensions that we can assess the results relative to the original data.

First, we perform three independent simulations along the following asymmetric double well potential,

\begin{align*}
    U(x) &= x^4 - 2 x^2 + 0.5 x,
\end{align*}

\begin{figure}[t!]
\centering
\includegraphics[width=0.5\textwidth]{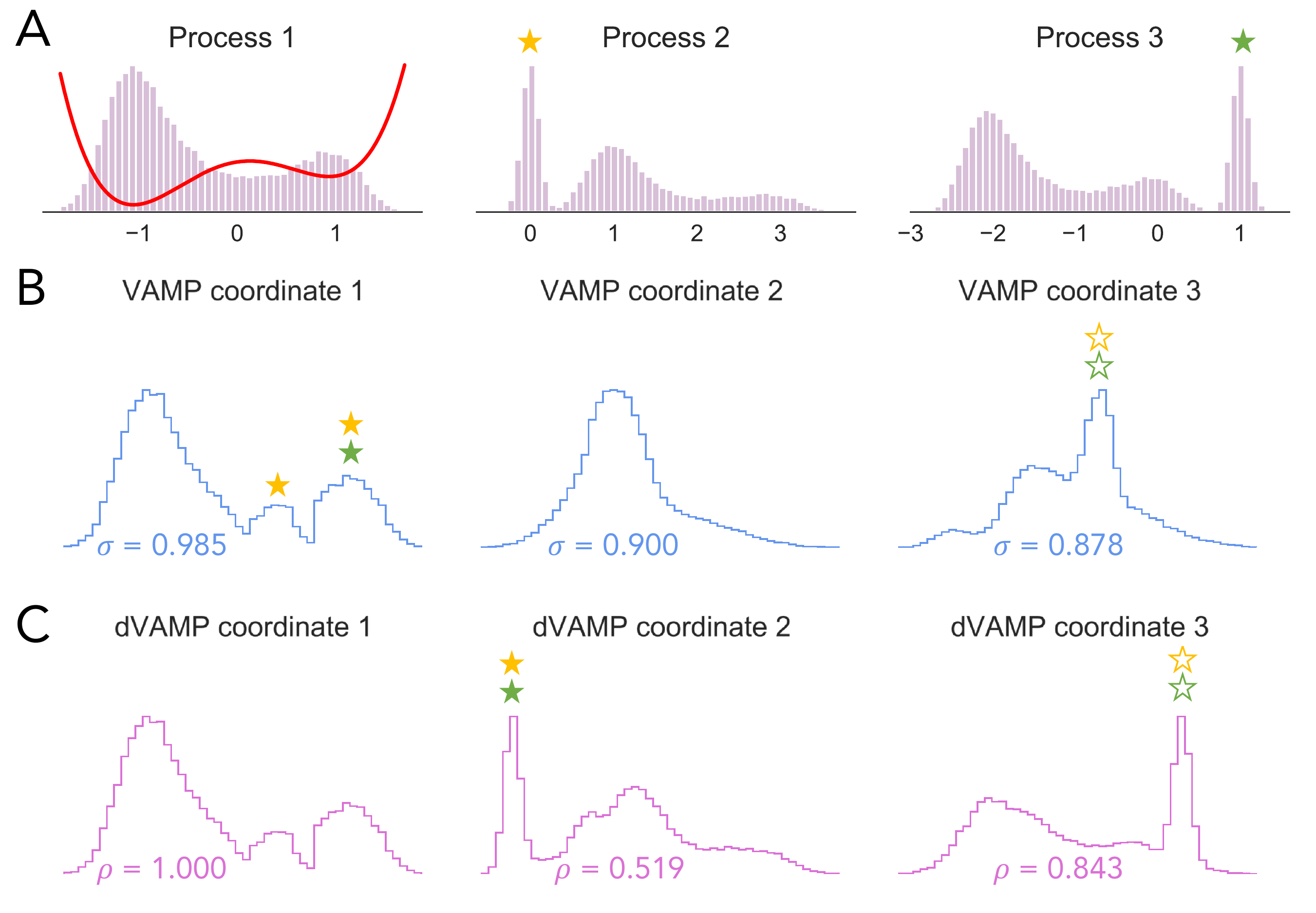}
\caption{
(A)~Histograms of particle distributions for the three processes in the toy dataset.
The potential used to simulate the first process is overlaid on the leftmost plot in red.
(B)~All histograms for the three reaction coordinates (transformations) produced from analysis with VAMP and (C)~dVAMP.
Gold and green stars indicate regions of the coordinates characterized by particle positions in the corresponding locations of the second and third processes, respectively.
Filled stars indicate that the region of the reaction coordinate is exclusively populated by those regions of the original processes, and unfilled stars indicate that other regions of the process can also be found at that location on the reaction coordinate.
The singular values $\sigma$ in~(B) are the same for the VAMP and dVAMP coordinates.
The $\rho$ values in~(C) are the Spearman rank correlation coefficients~\cite{spearman1904proof} of the dVAMP transformation with the corresponding VAMP transformation.
}
\label{fig:dw}
\end{figure}

\begin{figure*}[t!]
\centering
\includegraphics[width=\textwidth]{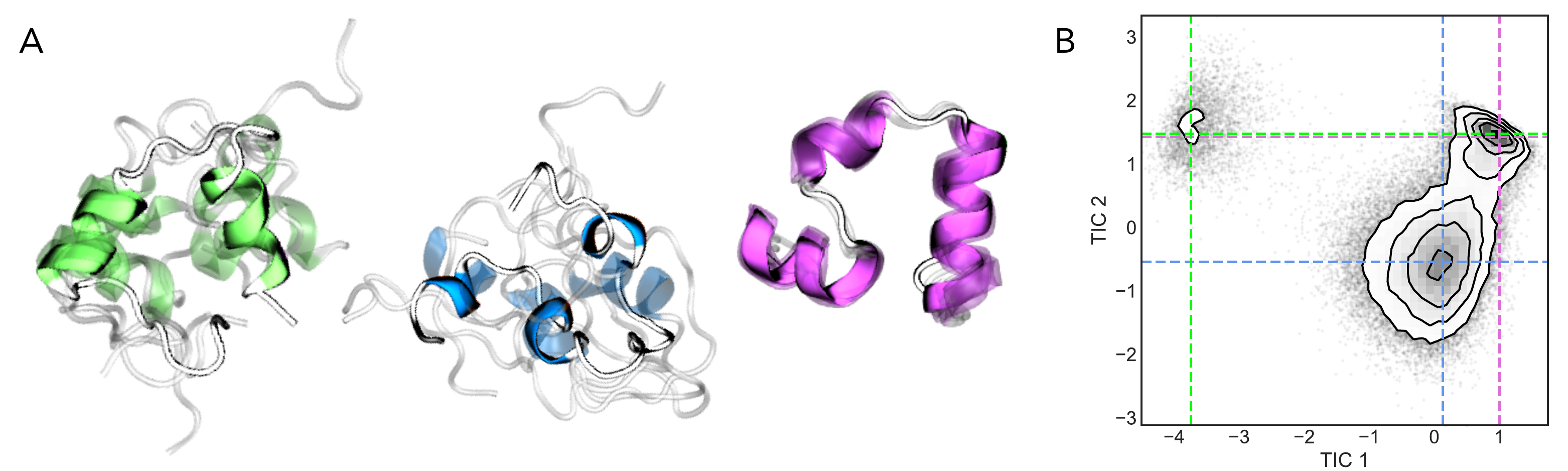}
\caption{
(A)~Sampled villin structures from the MD trajectory analyzed.
Helical secondary structure is colored and coils are white.
Each image represents five structures sampled from similar locations in TIC space as determined by a 250-center $k$-means model built upon the first three original TICs.
The purple structure represents the folded state, and the blue structure represents the denatured state.
The green structure is a rare helical misfolded state that we assert is an artifact.
(B)~Two-dimensional histograms for TICA transformations constructed from villin contact distances.
Dashed lines indicate the regions corresponding to the sampled structures of the same color.
The first TIC tracks the conversion to and from the rare artifact only.
The second TIC tracks the majority of the folding process and correlates well with RMSD to the folded structure.
}
\label{fig:villin}
\end{figure*}

\noindent{}which is visualized with a red line in the left plot of Fig.~\ref{fig:dw}A.
The first simulation produces the histogram of particle positions $P_1(n)$ for $n$ up to $50000$ simulation steps.
The histogram is illustrated on the same plot, and represents the first dimension of our toy dataset.

The other two independent simulations yield particle positions $P'_2(n)$ and $P'_3(n)$.
We then generate the second and third dimensions of our toy dataset from $P'_2$ and $P'_3$ so that they are dependent upon the first dimension $P_1$:

\begin{align*}
    P_2(n) &= \begin{cases}
    x \sim \mathcal{N}(0, 0.1^2) \text{, } &P_1(n) > 0  \\
    P'_2(n) + 2, &\text{otherwise}
    \end{cases} \\
    P_3(n) &= \begin{cases}
    x \sim \mathcal{N}(1, 0.1^2), &P_1(n) > 0.5 \\
    P'_3(n) - 1, &\text{otherwise},
    \end{cases}
\end{align*}

\noindent{}where $\mathcal{N}$ is the normal distribution.
The histograms of particle positions $P_2(n)$ and $P_3(n)$ are shown in the center and right plots of Fig.~\ref{fig:dw}A.

Now we wish to compare the results of the standard VAMP algorithm with the results of dVAMP.
To do this, we input the mean-free three-dimensional coordinates into Algs.~\ref{alg:vamp} and~\ref{alg:ccanipals} and compute the VAMP and dVAMP reaction coordinates.
Figures~\ref{fig:dw}B and~C illustrate the toy system results.
The first reaction coordinate is identical to the first deflated reaction coordinate because no deflation has been performed.
The singular values of the coordinates, provided in Fig.~\ref{fig:dw}B, are identical.

The difference between VAMP and dVAMP becomes apparent in the nondominant reaction coordinates.
The second VAMP coordinate is roughly Gaussian;
in contrast, the second dVAMP coordinate reveals meaningful structures within the original data.
Similarly, the third dVAMP transformation presents a more familiar structure with respect to the original data, whereas the third VAMP transformation is still ``contaminated'' by a Gaussian near the center of the coordinate.
The Spearman rank correlation coefficients~\cite{spearman1904proof} between pairs of corresponding reaction coordinates are also provided in Fig.~\ref{fig:dw}C, and the coefficients' distances from $1$ are consistent with qualitative assessments of their differences in structure.

\subsection{Folding of villin} \label{subsec:villin}

\subsubsection{Modeling with TICA}

Now we are interested in exploring how deflation can be used to tackle setbacks to kinetic modeling in a higher-dimensional protein example.
To do this, we analyze an MD simulation of the $35$-residue villin headpiece (PDB ID: 2f4k) performed by~\citet{lindorff2011fast}.
The folded state is depicted in the purple structures in Fig.~\ref{fig:villin}A.
The simulation dataset consists of a single $125$~$\mu$s trajectory with $34$ instances each of folding and unfolding.
For our analysis, we subsample the original dataset by a factor of $10$ to produce $2$~ns timesteps.

Before constructing a kinetic model, we convert the Cartesian coordinates output from the simulation into features; in this case, a set of pairwise contact distances, where the distance is defined by the closest heavy atoms in the pair.
For our analysis, we consider the pairwise distances among a subset of twelve residues at three-residue intervals starting with the first (LEU~1, GLU~4, \dots, LEU~34) for a total of $66$ distances.

Now, we want to use TICA to build a kinetic model for our data.
To perform TICA, we can use VAMP (Alg.~\ref{alg:vamp}) with the reversibility conditions in Sec.~\ref{sec:tica}.
This is achieved by letting $\mathbf{X} \gets \mathbf{X'}^\frown\mathbf{Y'}$ and $\mathbf{Y} \gets  \mathbf{Y'}^\frown\mathbf{X'}$,
where $\mathbf{X}'$ and $\mathbf{Y}'$ are the original time-lagged data and~$^\frown$ signifies concatenation along the row axis.
We use a lag time of $\tau = 50$~ns, or $25$ time steps.
The TICA results are shown in a two-dimensional histogram in Fig.~\ref{fig:villin}B.
The first TICA coordinate separates the majority of configuration space from a rare misfolded state (green in Fig.~\ref{fig:villin}A).
The second TICA coordinate separates the denatured state from the folded state (blue and purple in Fig.~\ref{fig:villin}A, respectively).

In order to demonstrate the application of dVAMP and subsequent analysis with deflation, we will now assert that we are \emph{not} interested in the rare misfolded state due to its poor sampling, and that it is therefore an artifact that is disruptive to our kinetic model.
If we wanted to stop at TICA, we could ignore the first TIC.
However, we may be interested in any of the following further analyses:
\begin{enumerate}[label=(\roman*)]
    \item creating an MSM~\cite{husic2018markov}, especially when using kinetic mapping~\cite{noe2015kinetic} (in which the slowest modes are most important to the MSM state decomposition);
    \item evaluating the approximation quality of that MSM using the VAC~\cite{noe2013variational,mcgibbon2015variational};
    \item macrostating our TICA model or MSM using an eigenvector-based approach such as Perron cluster cluster analysis (PCCA)~\cite{deuflhard2000identification};
    \item constructing a VAMPnet~\cite{mardt2018vampnets} to obtain a kinetic model (similar to performing (i)-(iii) in sequence);
\end{enumerate}
or the application of a variety of other tools.

What (i)-(iv) have in common is that if an undesired slow process is present in the model, then it will correspondingly affect analysis in the following ways:
\begin{enumerate}[label=(\roman*)]
    \item the MSM state decomposition will be influenced by the artifact,
    \item MSM variational scores will be a function of how accurately the artifact kinetics are modeled,
    \item divisive clustering methods such as PCCA will divide macrostates according to the artifact, and
    \item the VAMPnet optimization will involve optimally modeling the artifact. 
\end{enumerate}

\subsubsection{Removing undesired processes with deflation}

The state of the art in Markovian kinetic modeling has not yet provided a solution to these problems other than the acquiescence that if we are approximating system kinetics then we must work with whatever the global kinetics are.
However, the use of deflation provides a new option.
In Sec.~\ref{subsec:dw}, we saw how the use of dVAMP can reveal structure in nondominant reaction coordinates.

Now, we will \emph{selectively} use deflation to prevent our model from optimizing the kinetics of the precise process we do not want.
Rewriting line 11 in Alg.~\ref{alg:ccanipals} for this purpose, we have,

\begin{align}
    \mathbf{X}_\text{new} = \mathbf{X}_\text{original} - \xi_c[\mathbf{X}_\text{original}^\top\mathbf{\xi}_c(\mathbf{\xi}_c^\top\mathbf{\xi}_c)^{-1}]^\top, \label{eq:deflation}
\end{align}

\noindent{}where $c$ is the index of the coordinate that is undesired, and $\mathbf{\xi}_c$ is the \emph{score} of that coordinate, which was calculated in the course of Alg.~\ref{alg:ccanipals}.\footnote{This is different from $\mathbf{\xi}'_c$ in Eqn.~\ref{eq:scorex}, which is not calculated in the course of Alg.~\ref{alg:ccanipals} on data that has been deflated $c$ times, but instead is calculated on the original data afterward.}
This removes the effect of the $c$th coordinate on the dynamics by setting the eigen- or singular value of the associated eigen- or singular vector of the matrix product $\tilde{\mathbf{K}}$~\eqref{eq:ktilde} to zero.

For our purposes, this means that deflation has rendered the corresponding dynamical process \emph{fast}:~it appears as noise in the new basis.
Because the VAC, VAMP, and related methods seek to optimally approximate the slowest processes in the system, deflation effectively ``removes'' this process from the optimization procedure of the global kinetic model by lowering its timescale in the new feature basis.

To achieve this for villin, we perform dVAMP as in Sec.~\ref{subsec:dw} through the first dVAMP coordinate in order to obtain the scores for the undesired (slowest) process.
Then, we apply~\eqref{eq:deflation} with $c=1$.\footnote{
Note that if the undesired process is not the slowest process, i.e., $c > 1$, we can deflate it exclusively using~\eqref{eq:deflation} once for $c$ only. However, we must perform dVAMP \emph{through} the $c$th coordinates to obtain the necessary weights.}
Recall that our data $\mathbf{X}_\text{original}$ are the $66$ pairwise contact distances, and so $\mathbf{X}_\text{new}$ are corresponding \emph{deflated} distances.\footnote{To calculate all of the distances, we perform the same transformation for $\mathbf{Y}$ using $\omega_c$ in order to obtain the last $\tau$ data points, which are not present in $\mathbf{X}$ due to the time lag.}
The deflated distances form a basis in which the eigenvalue of an undesired process is set to zero.
They can now be input into any model ranking framework (e.g., MSMs or VAMPnets with variable hyperparameters) such that the resulting kinetic models will not be optimizing the artifact kinetics alongside the interesting ones.

\subsubsection{Constructing VAMPnets for villin folding}

\begin{figure}[t!]
\centering
\includegraphics[width=0.45\textwidth]{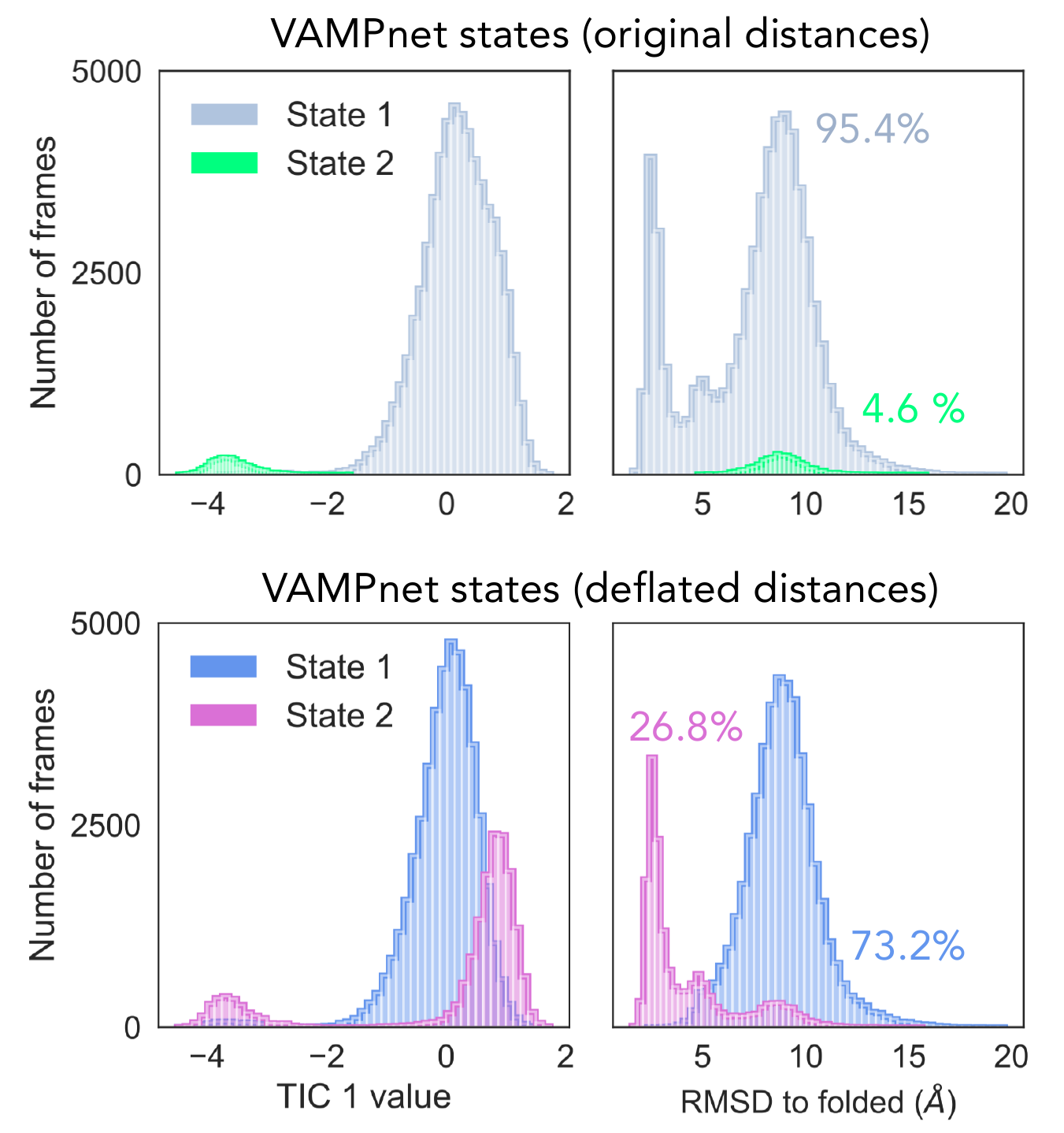}
\caption{
One-dimensional histograms for two-state VAMPnets constructed from the original distances (top row) and deflated distances (bottom row).
Each row shows two different analyses of the same data.
The use of two states was chosen as a hyperparameter \emph{a priori}.
The hard state assignments are determined by mapping soft state memberships to the state with the higher probability.
The first VAMPnet (original distances) separates the rare artifact from the rest of the configurations.
The second VAMPnet (deflated distances) separates structures near the folded state from the denatured state.
Percentages indicate state populations.
}
\label{fig:vampnets}
\end{figure}

We now investigate the difference in modeling results between the original and deflated bases using VAMPnets~\cite{mardt2018vampnets}.
VAMP (Alg.~\ref{alg:vamp}) calculates an SVD in line 1 which produces a diagonal $k \times k$ matrix $\mathbf{S}$ of positive singular values $\sigma_i, i \leq k$.
The sum of the highest $k$ singular values raised to the $r$th power was proven by~\citet{wu2017vamp} to be a variational score---the VAMP-$r$ score---for the approximation quality of the kinetic model:

\begin{align}
    \text{VAMP-}r \equiv \sum_{i=1}^k \sigma_i^r.
\end{align}

It follows from the existence of a scalar variational bound that machine learning algorithms can be designed to optimize it~\cite{mcgibbon2015variational}.
Using neural networks, VAMPnets achieve exactly this by constructing few-state Koopman models directly from features by optimizing a VAMP score~\cite{mardt2018vampnets}.
Here, we build VAMPnets for both our original (mean-free) contact distances and our new, deflated basis.
We chose VAMPnets because, due to the neural network architecture, there is no opportunity to manually modify a model while it is being built (e.g., by manually removing an undesired TIC).

For our VAMPnets, we use a lag time of $300$~ns and choose two metastable states.
The training and validation sets are split to be approximately equal in size ($51$\% and $49$\% of the data, respectively), and $10$ epochs are performed when training each model.
The remaining hyperparameters are the defaults at the time of modeling (batch size, 1000; network depth, 6; layer width, 100; learning rate, 0.0001).
The VAMPnet output is a list of state assignment probabilities for each frame in the trajectory; binary (``hard'') memberships are determined by assigning each frame to the state with the greater probability.

Our VAMPnet results are illustrated in Fig.~\ref{fig:vampnets}.
The top row of histograms summarizes the VAMPnet constructed on the original contact distances.
The upper left plot shows distributions for each (hard) VAMPnet state across the values of the first TIC from the original TICA analysis.
The upper right plot shows distributions for the same states across root-mean-square deviation (RMSD) values to the folded structure of villin.
It is clear from these plots that the two-state VAMPnet separates our artifact (4.6\% of frames) from the rest of the ensemble.

The second row of histograms shows the same analysis for the VAMPnet constructed on the deflated contact distances with the same hyperparameters.
The lower left plot shows that both VAMPnet states contain structures from the artifact region.
The lower right plot shows that the new two-state VAMPnet generally isolates the folded structure (26.8\%) from the denatured ensemble.
Thus, by using deflated contact distances, we obtain a VAMPnet that captures folding and unfolding using the same parameters with which we were unable to capture folding and unfolding before the basis deflation.

We can see from the lower left histogram in Fig.~\ref{fig:vampnets} that the ``artifact'' is present in the folded state, and further analysis shows that the majority of structures assigned to the folded state with RMSDs greater than $7.5$~\AA~to the folded structure are ``artifact'' structures (as defined by their location along the first TIC).
We do not claim that this analysis---a two-state VAMPnet with deflated distances---is the best one for villin (for example, previous research suggests a three-state model is better for describing the same dataset~\cite{beauchamp2012simple}).
Instead, we have demonstrated that dVAMP can be used to prevent a kinetic model from optimizing the accuracy of a (subjectively) undesired process \emph{a priori}.


\section{Discussion}

We have examined incorporation of deflation into state-of-the-art methods in molecular kinetics.
It is already known that VAMP is equivalent to CCA without deflation.
In this study we explore the use of dVAMP---VAMP with deflation---for the development of reaction coordinates.
Then, we demonstrate the selective use of deflation to remove a specific slow process from a kinetic model; in this case, a VAMPnet.

In the toy example in Sec.~\ref{subsec:dw}, we see that nondominant reaction coordinates obtained from dVAMP qualitatively provide more information about the system than their nondeflated VAMP counterparts.
In addition to increased interpretability, these coordinates might be used for enhanced sampling.
For example, MD simulations can be started from the top of a barrier revealed only in a deflated reaction coordinate.

In the protein folding example, we see that deflation can obscure a long-timescale process from a kinetic model in order to facilitate further analysis that is desired to be independent of that process.
In MD simulations, the dominant processes may not be the same as the processes of interest due to low sampling or force field artifacts~\cite{banushkina2015nonparametric,mcgibbon2017identification}.
Thus, deflation presents a systematic way of removing the effects of these undesired modes on model building.
Using deflated data, MSMs, Koopman models, or VAMPnets can be constructed such that the eigen- or singular vectors of the resulting model capture only those dynamics which are under investigation.
Furthermore, multiple models can be compared with the guarantee none of them are optimizing the accuracy of the undesired slow mode.

Frameworks such as MSMs and Koopman models require that the dynamical modes are orthogonal.
Thus, if dynamical modes of interest are faster than an undesired process, they are constrained by this orthogonality condition.
By deflating the basis, we gain the flexibility to optimally model the dynamical modes of interest without requiring that they are orthogonal to something uninteresting.

This work opens the door not only for improved analysis but also for related methods developments.
Often we are interested in some separate \textit{observable} that changes in time with the data and its relationship to the features.
The implementation of a deflation method in the context of explaining the relationships between a feature and an observable is a valuable pursuit and has already been explored in the time-independent context~\cite{hub2009detection,krivobokova2012partial}.
The adaptation of such methods to time-series dynamics would be of great utility to the field and the advance presented here may serve as a starting point.

Several open source Python software packages facilitated this study.
The double well potential in Sec.~\ref{subsec:dw} was simulated using PyEMMA~\cite{wehmeyer2018introduction}.
The 2D histogram in Fig.~\ref{fig:villin} was generated using Corner~\cite{foreman2016corner}, contact distances for villin were calculated with MDTraj~\cite{mcgibbon2015mdtraj}, VAMPnets were calculated using the deeptime package~\cite{mardt2018vampnets}.
An example Jupyter notebook~\cite{jupyter} is provided to reproduce the analysis in Sec.~\ref{subsec:dw} which also uses NumPy~\cite{numpy}, SciPy~\cite{scipy}, Scikit-learn~\cite{scikit-learn}, Matplotlib~\cite{matplotlib}, and Seaborn~\cite{seaborn}.
Protein trajectories were visualized with VMD~\cite{vmd}.




\section*{Acknowledgements}

BEH is extremely grateful to Fabian Paul, Muneeb Sultan, Moritz Hoffmann, Andreas Mardt, Luca Pasquali, Stefan Klus, Tim Hempel, and Chris Wehmeyer for helpful discussions and feedback.
We thank D.~E.~Shaw Research for providing the protein simulation dataset.
BEH and FN acknowledge funding from the European Commission (ERC CoG 772230 ``ScaleCell'') and the Deutsche Forschungsgemeinschaft (SFB1114/A04).


\appendix

\section{Intuition for the relationship between VAMP and canonical CCA} \label{app:intuition}

We can demonstrate the equivalence of VAMP and canonical CCA (``dVAMP'' for time-lagged data) by considering the calculation of the first component, which is identical in both formulations.

By making substitutions from lines 4--8 of Alg.~\ref{alg:ccanipals}, we can write,

\begin{align}
    \phi_c &\leftarrow \mathbf{X}^{+}\mathbf{Y}\mathbf{Y}^{+}\mathbf{X}\phi_c.
\end{align}

\noindent{}Further substituting the definition of the Moore-Penrose pseudoinverse~\cite{penrose1955generalized}, $\mathbf{X}^{+} \equiv (\mathbf{X}^\top\mathbf{X})^{-1}\mathbf{X}^\top$, we obtain,

\begin{align}
    \phi_c &\leftarrow (\mathbf{X}^\top\mathbf{X})^{-1}\mathbf{X}^\top\mathbf{Y}(\mathbf{Y}^\top\mathbf{Y})^{-1}\mathbf{Y}^\top\mathbf{X}\phi_c,
\end{align}

\noindent{}which has been shown to be equivalent to the power method for identifying the dominant eigenvector of a matrix~\cite{wegelin2000survey}.
Thus we see that $\phi_c$ is the dominant eigenvector of the expression,

\begin{align}
    (\mathbf{X}^\top\mathbf{X})^{-1}(\mathbf{X}^\top\mathbf{Y})(\mathbf{Y}^\top\mathbf{Y})^{-1}(\mathbf{Y}^\top\mathbf{X}), \label{eq:cca_eig}
\end{align}

\noindent{}for $\mathbf{X}$ at a given iteration (i.e., possibly deflated).

To obtain an intuition for the connection of this expression~\eqref{eq:cca_eig} to the VAMP solution in Alg.~\ref{alg:vamp}, we first substitute~\eqref{eq:xtilde},~\eqref{eq:ytilde}, and~\eqref{eq:ktilde} into $\tilde{\mathbf{Y}} = \tilde{\mathbf{X}}\tilde{\mathbf{K}}$,

\begin{align}
    \mathbf{Y}{\mathbf{C}}_{\tau\tau}^{-\frac{1}{2}} &=  (\mathbf{X}{\mathbf{C}}_{00}^{-\frac{1}{2}}) (\mathbf{C}_{00}^{-\frac{1}{2}}\mathbf{C}_{0\tau}\mathbf{C}_{\tau\tau}^{-\frac{1}{2}}),
\end{align}

\noindent{}which simplifies to,

\begin{align}
    \mathbf{Y} &= \mathbf{X} \mathbf{C}_{00}^{-1}\mathbf{C}_{0\tau}.
\end{align}

\noindent{}Thus the matrix that converts $\mathbf{X}$ to $\mathbf{Y}$ (i.e., propagates the time-series if $\mathbf{X}$ and $\mathbf{Y}$ represent time-lagged data) is,

\begin{align}
    \mathbf{K}_f &\equiv \mathbf{C}_{00}^{-1}\mathbf{C}_{0\tau} = (\mathbf{X}^\top\mathbf{X})^{-1}\mathbf{X}^\top\mathbf{Y},
\end{align}

\noindent{}where the subscript $f$ indicates that, in the time-lagged interpretation, the matrix propagates the data \emph{forward} in time.
It is easy to see that for the backward direction,

\begin{align}
    \mathbf{K}_b &\equiv \mathbf{C}_{\tau\tau}^{-1}\mathbf{C}_{0\tau}^\top = (\mathbf{Y}^\top\mathbf{Y})^{-1}\mathbf{Y}^\top\mathbf{X}.
\end{align}

Now, we can see that~\eqref{eq:cca_eig} is equal to $\mathbf{K}_f\mathbf{K}_b$, as well as the analogous solution for the $\mathbf{Y}$ weights:

\begin{align}
    (\mathbf{X}^\top\mathbf{X})^{-1}(\mathbf{X}^\top\mathbf{Y})(\mathbf{Y}^\top\mathbf{Y})^{-1}(\mathbf{Y}^\top\mathbf{X}) &= \mathbf{K}_f\mathbf{K}_b \label{eq:kfkb}\\
    (\mathbf{Y}^\top\mathbf{Y})^{-1}(\mathbf{Y}^\top\mathbf{X})(\mathbf{X}^\top\mathbf{X})^{-1}(\mathbf{X}^\top\mathbf{Y}) &= \mathbf{K}_b\mathbf{K}_f. \label{eq:kbkf}
\end{align}

\noindent{}CCA with deflation via NIPALS obtains the dominant eigenvector of the left-hand sides, whereas VAMP calculates the dominant eigenvector of the right-hand sides.
The dominant eigenvalues of the expressions in~\eqref{eq:kfkb} and~\eqref{eq:kbkf} are the squares of the dominant singular values obtained from VAMP and dVAMP.

In general, the equivalences above apply when $\mathbf{X}$ and $\mathbf{Y}$ are the same for both algorithms, and only the dominant component is considered.

\section{CCA with iterative deflation via the SVD} \label{app:ccasvd}

Algorithm~\ref{alg:ccanipals} can be rewritten with an SVD in place of the inner NIPALS loop as follows:

\begin{algorithm}[H]
\begin{algorithmic}[1]
\Require $\mathbf{X} \in \mathbb{R}^{n \times m}, \mathbf{Y} \in \mathbb{R}^{n \times p}$\Comment{mean-free columns}
\For{each component $c$}
\State $\mathbf{U'SV'}^\top \gets (\mathbf{X}^\top\mathbf{X})^{-\frac{1}{2}} \mathbf{X}^\top\mathbf{Y} (\mathbf{Y}^\top\mathbf{Y})^{-\frac{1}{2}}$
\State $\mathbf{U} \gets n^{\frac{1}{2}}(\mathbf{X}^\top\mathbf{X})^{-\frac{1}{2}} \mathbf{U}' $
\State $\mathbf{V} \gets n^{\frac{1}{2}} (\mathbf{Y}^\top\mathbf{Y})^{-\frac{1}{2}} \mathbf{V}' $
\State $\mathbf{\phi}_c \gets$ first column of $\mathbf{U}$ \Comment{$\mathbf{\phi}_c \equiv c$th \textit{x weight} vector}
\State $\mathbf{\psi}_c \gets$ first column of $\mathbf{V}$ \Comment{$\mathbf{\psi}_c \equiv c$th \textit{y weight} vector}\\
\State $\mathbf{\xi}_c \gets \mathbf{X}\mathbf{\phi}_c$  \Comment{$\mathbf{\xi}_c \equiv c$th \textit{x score} vector}
\State $\mathbf{\omega}_c \gets \mathbf{Y}\mathbf{\psi}_c$ \Comment{$\mathbf{\omega}_c \equiv c$th \textit{y score} vector} \\
\State $\mathbf{X} \gets \mathbf{X} - \xi_c[\mathbf{X}^\top\mathbf{\xi}_c(\mathbf{\xi}_c^\top\mathbf{\xi}_c)^{-1}]^\top$ \Comment{Deflation of $\mathbf{X}$}
\State $\mathbf{Y} \gets \mathbf{Y} - \omega_c[\mathbf{Y}^\top\mathbf{\omega}_c(\mathbf{\omega}_c^\top\mathbf{\omega}_c)^{-1}]^\top$ \Comment{Deflation of $\mathbf{Y}$}
\EndFor
\caption{\label{alg:ccasvd} CCA (SVD with iterative deflation)}
\end{algorithmic}
\end{algorithm}

However, the implementation of this algorithm contains multiple inverse square roots, which will be numerically unstable with deflation, leading to errors that are propagated with each new component.
Regularization can be employed to avoid ill-conditioned covariance matrices, but then the solution should no longer be expected to match that of the algorithm with NIPALS.
Indeed, it is generally a good idea to regularize covariance matrices, including in the NIPALS algorithm and TICA.
The interested reader is referred to Appendix B in Ref.~\onlinecite{mcgibbon2017identification}.

\section{Connection to Koopman operator approximation} \label{app:operator}

The matrix $\mathbf{K}$ is a finite-dimensional linear approximation to the continuous integral Koopman operator $\mathcal{K}$ and can be called the Koopman matrix~\cite{wu2017variational,wu2017vamp,paul2019identification,koopman1931hamiltonian,mezic2005spectral,klus2018data}.
In the time-lagged context, the operator propagates a view of the system at time $t$ to a possibly different view of the system at time $t+\tau$.
While we have focused on obtaining reaction coordinates, it is useful to note that the weight (singular) \emph{vectors} obtained in VAMP (Alg.~\ref{alg:vamp}) are approximations to the singular \emph{functions} of the Koopman operator.
The first weight vector calculation in dVAMP (Alg.~\ref{alg:ccanipals}) approximates the dominant singular functions because it is equivalent to the single SVD calculation in VAMP (Alg.~\ref{alg:vamp}; see Appendices~\ref{app:intuition} and~
\ref{app:ccasvd}).

However, if a relationship between the subsequent weight vectors calculated in dVAMP and $\mathcal{K}$ exists, it is not straightforward.
In a sense, a new operator is approximated with each deflation for a matrix built from new, deflated datasets $\mathbf{X}_c$ and $\mathbf{Y}_c$, in which the presence of the previously calculated $c$ weight vectors have been removed.
Thus, each new set of weight vectors approximates the dominant singular functions of some \emph{new} operator $\mathcal{K}_c$, which describes the deflated dynamics.

\bibliography{refs}

\begin{thebibliography}{51}%
\makeatletter
\providecommand \@ifxundefined [1]{%
 \@ifx{#1\undefined}
}%
\providecommand \@ifnum [1]{%
 \ifnum #1\expandafter \@firstoftwo
 \else \expandafter \@secondoftwo
 \fi
}%
\providecommand \@ifx [1]{%
 \ifx #1\expandafter \@firstoftwo
 \else \expandafter \@secondoftwo
 \fi
}%
\providecommand \natexlab [1]{#1}%
\providecommand \enquote  [1]{``#1''}%
\providecommand \bibnamefont  [1]{#1}%
\providecommand \bibfnamefont [1]{#1}%
\providecommand \citenamefont [1]{#1}%
\providecommand \href@noop [0]{\@secondoftwo}%
\providecommand \href [0]{\begingroup \@sanitize@url \@href}%
\providecommand \@href[1]{\@@startlink{#1}\@@href}%
\providecommand \@@href[1]{\endgroup#1\@@endlink}%
\providecommand \@sanitize@url [0]{\catcode `\\12\catcode `\$12\catcode
  `\&12\catcode `\#12\catcode `\^12\catcode `\_12\catcode `\%12\relax}%
\providecommand \@@startlink[1]{}%
\providecommand \@@endlink[0]{}%
\providecommand \url  [0]{\begingroup\@sanitize@url \@url }%
\providecommand \@url [1]{\endgroup\@href {#1}{\urlprefix }}%
\providecommand \urlprefix  [0]{URL }%
\providecommand \Eprint [0]{\href }%
\providecommand \doibase [0]{http://dx.doi.org/}%
\providecommand \selectlanguage [0]{\@gobble}%
\providecommand \bibinfo  [0]{\@secondoftwo}%
\providecommand \bibfield  [0]{\@secondoftwo}%
\providecommand \translation [1]{[#1]}%
\providecommand \BibitemOpen [0]{}%
\providecommand \bibitemStop [0]{}%
\providecommand \bibitemNoStop [0]{.\EOS\space}%
\providecommand \EOS [0]{\spacefactor3000\relax}%
\providecommand \BibitemShut  [1]{\csname bibitem#1\endcsname}%
\let\auto@bib@innerbib\@empty
\bibitem [{\citenamefont {Husic}\ and\ \citenamefont
  {Pande}(2018)}]{husic2018markov}%
  \BibitemOpen
  \bibfield  {author} {\bibinfo {author} {\bibfnamefont {B.~E.}\ \bibnamefont
  {Husic}}\ and\ \bibinfo {author} {\bibfnamefont {V.~S.}\ \bibnamefont
  {Pande}},\ }\bibfield  {title} {\enquote {\bibinfo {title} {Markov state
  models: From an art to a science},}\ }\href@noop {} {\bibfield  {journal}
  {\bibinfo  {journal} {J. Am. Chem. Soc.}\ }\textbf {\bibinfo {volume}
  {140}},\ \bibinfo {pages} {2386--2396} (\bibinfo {year} {2018})}\BibitemShut
  {NoStop}%
\bibitem [{\citenamefont {Sch{\"u}tte}\ and\ \citenamefont
  {Sarich}(2015)}]{schutte2015critical}%
  \BibitemOpen
  \bibfield  {author} {\bibinfo {author} {\bibfnamefont {C.}~\bibnamefont
  {Sch{\"u}tte}}\ and\ \bibinfo {author} {\bibfnamefont {M.}~\bibnamefont
  {Sarich}},\ }\bibfield  {title} {\enquote {\bibinfo {title} {A critical
  appraisal of markov state models},}\ }\href@noop {} {\bibfield  {journal}
  {\bibinfo  {journal} {Eur. Phys. J.}\ }\textbf {\bibinfo {volume} {224}},\
  \bibinfo {pages} {2445--2462} (\bibinfo {year} {2015})}\BibitemShut {NoStop}%
\bibitem [{\citenamefont {No{\'e}}\ and\ \citenamefont
  {N{\"u}ske}(2013)}]{noe2013variational}%
  \BibitemOpen
  \bibfield  {author} {\bibinfo {author} {\bibfnamefont {F.}~\bibnamefont
  {No{\'e}}}\ and\ \bibinfo {author} {\bibfnamefont {F.}~\bibnamefont
  {N{\"u}ske}},\ }\bibfield  {title} {\enquote {\bibinfo {title} {A variational
  approach to modeling slow processes in stochastic dynamical systems},}\
  }\href@noop {} {\bibfield  {journal} {\bibinfo  {journal} {Multiscale Model.
  Simul.}\ }\textbf {\bibinfo {volume} {11}},\ \bibinfo {pages} {635--655}
  (\bibinfo {year} {2013})}\BibitemShut {NoStop}%
\bibitem [{\citenamefont {Wu}\ and\ \citenamefont
  {No{\'e}}(2017)}]{wu2017vamp}%
  \BibitemOpen
  \bibfield  {author} {\bibinfo {author} {\bibfnamefont {H.}~\bibnamefont
  {Wu}}\ and\ \bibinfo {author} {\bibfnamefont {F.}~\bibnamefont {No{\'e}}},\
  }\bibfield  {title} {\enquote {\bibinfo {title} {Variational approach for
  learning {Markov} processes from time series data},}\ }\href@noop {}
  {\bibfield  {journal} {\bibinfo  {journal} {arXiv preprint arXiv:1707.04659}\
  } (\bibinfo {year} {2017})}\BibitemShut {NoStop}%
\bibitem [{\citenamefont {Wu}\ \emph {et~al.}(2017)\citenamefont {Wu},
  \citenamefont {N{\"u}ske}, \citenamefont {Paul}, \citenamefont {Klus},
  \citenamefont {Koltai},\ and\ \citenamefont {No{\'e}}}]{wu2017variational}%
  \BibitemOpen
  \bibfield  {author} {\bibinfo {author} {\bibfnamefont {H.}~\bibnamefont
  {Wu}}, \bibinfo {author} {\bibfnamefont {F.}~\bibnamefont {N{\"u}ske}},
  \bibinfo {author} {\bibfnamefont {F.}~\bibnamefont {Paul}}, \bibinfo {author}
  {\bibfnamefont {S.}~\bibnamefont {Klus}}, \bibinfo {author} {\bibfnamefont
  {P.}~\bibnamefont {Koltai}}, \ and\ \bibinfo {author} {\bibfnamefont
  {F.}~\bibnamefont {No{\'e}}},\ }\bibfield  {title} {\enquote {\bibinfo
  {title} {Variational koopman models: slow collective variables and molecular
  kinetics from short off-equilibrium simulations},}\ }\href@noop {} {\bibfield
   {journal} {\bibinfo  {journal} {J. Chem. Phys.}\ }\textbf {\bibinfo {volume}
  {146}},\ \bibinfo {pages} {154104} (\bibinfo {year} {2017})}\BibitemShut
  {NoStop}%
\bibitem [{\citenamefont {Mardt}\ \emph {et~al.}(2018)\citenamefont {Mardt},
  \citenamefont {Pasquali}, \citenamefont {Wu},\ and\ \citenamefont
  {No{\'e}}}]{mardt2018vampnets}%
  \BibitemOpen
  \bibfield  {author} {\bibinfo {author} {\bibfnamefont {A.}~\bibnamefont
  {Mardt}}, \bibinfo {author} {\bibfnamefont {L.}~\bibnamefont {Pasquali}},
  \bibinfo {author} {\bibfnamefont {H.}~\bibnamefont {Wu}}, \ and\ \bibinfo
  {author} {\bibfnamefont {F.}~\bibnamefont {No{\'e}}},\ }\bibfield  {title}
  {\enquote {\bibinfo {title} {Vampnets for deep learning of molecular
  kinetics},}\ }\href@noop {} {\bibfield  {journal} {\bibinfo  {journal}
  {Nature Commun.}\ }\textbf {\bibinfo {volume} {9}},\ \bibinfo {pages} {5}
  (\bibinfo {year} {2018})}\BibitemShut {NoStop}%
\bibitem [{\citenamefont {Scherer}\ \emph {et~al.}(2019)\citenamefont
  {Scherer}, \citenamefont {Husic}, \citenamefont {Hoffmann}, \citenamefont
  {Paul}, \citenamefont {Wu},\ and\ \citenamefont
  {No{\'e}}}]{scherer2019variational}%
  \BibitemOpen
  \bibfield  {author} {\bibinfo {author} {\bibfnamefont {M.~K.}\ \bibnamefont
  {Scherer}}, \bibinfo {author} {\bibfnamefont {B.~E.}\ \bibnamefont {Husic}},
  \bibinfo {author} {\bibfnamefont {M.}~\bibnamefont {Hoffmann}}, \bibinfo
  {author} {\bibfnamefont {F.}~\bibnamefont {Paul}}, \bibinfo {author}
  {\bibfnamefont {H.}~\bibnamefont {Wu}}, \ and\ \bibinfo {author}
  {\bibfnamefont {F.}~\bibnamefont {No{\'e}}},\ }\bibfield  {title} {\enquote
  {\bibinfo {title} {Variational selection of features for molecular
  kinetics},}\ }\href@noop {} {\bibfield  {journal} {\bibinfo  {journal} {J.
  Chem. Phys.}\ }\textbf {\bibinfo {volume} {150}},\ \bibinfo {pages} {194108}
  (\bibinfo {year} {2019})}\BibitemShut {NoStop}%
\bibitem [{\citenamefont {Banushkina}\ and\ \citenamefont
  {Krivov}(2015)}]{banushkina2015nonparametric}%
  \BibitemOpen
  \bibfield  {author} {\bibinfo {author} {\bibfnamefont {P.~V.}\ \bibnamefont
  {Banushkina}}\ and\ \bibinfo {author} {\bibfnamefont {S.~V.}\ \bibnamefont
  {Krivov}},\ }\bibfield  {title} {\enquote {\bibinfo {title} {Nonparametric
  variational optimization of reaction coordinates},}\ }\href@noop {}
  {\bibfield  {journal} {\bibinfo  {journal} {J. Chem. Phys.}\ }\textbf
  {\bibinfo {volume} {143}},\ \bibinfo {pages} {184108} (\bibinfo {year}
  {2015})}\BibitemShut {NoStop}%
\bibitem [{\citenamefont {McGibbon}, \citenamefont {Husic},\ and\ \citenamefont
  {Pande}(2017)}]{mcgibbon2017identification}%
  \BibitemOpen
  \bibfield  {author} {\bibinfo {author} {\bibfnamefont {R.~T.}\ \bibnamefont
  {McGibbon}}, \bibinfo {author} {\bibfnamefont {B.~E.}\ \bibnamefont {Husic}},
  \ and\ \bibinfo {author} {\bibfnamefont {V.~S.}\ \bibnamefont {Pande}},\
  }\bibfield  {title} {\enquote {\bibinfo {title} {Identification of simple
  reaction coordinates from complex dynamics},}\ }\href@noop {} {\bibfield
  {journal} {\bibinfo  {journal} {J. Chem. Phys.}\ }\textbf {\bibinfo {volume}
  {146}},\ \bibinfo {pages} {044109} (\bibinfo {year} {2017})}\BibitemShut
  {NoStop}%
\bibitem [{\citenamefont {Naritomi}\ and\ \citenamefont
  {Fuchigami}(2011)}]{naritomi2011slow}%
  \BibitemOpen
  \bibfield  {author} {\bibinfo {author} {\bibfnamefont {Y.}~\bibnamefont
  {Naritomi}}\ and\ \bibinfo {author} {\bibfnamefont {S.}~\bibnamefont
  {Fuchigami}},\ }\bibfield  {title} {\enquote {\bibinfo {title} {Slow dynamics
  in protein fluctuations revealed by time-structure based independent
  component analysis: the case of domain motions},}\ }\href@noop {} {\bibfield
  {journal} {\bibinfo  {journal} {J. Chem. Phys.}\ }\textbf {\bibinfo {volume}
  {134}},\ \bibinfo {pages} {065101} (\bibinfo {year} {2011})}\BibitemShut
  {NoStop}%
\bibitem [{\citenamefont {Schwantes}\ and\ \citenamefont
  {Pande}(2013)}]{schwantes2013improvements}%
  \BibitemOpen
  \bibfield  {author} {\bibinfo {author} {\bibfnamefont {C.~R.}\ \bibnamefont
  {Schwantes}}\ and\ \bibinfo {author} {\bibfnamefont {V.~S.}\ \bibnamefont
  {Pande}},\ }\bibfield  {title} {\enquote {\bibinfo {title} {Improvements in
  {Markov} state model construction reveal many non-native interactions in the
  folding of {NTL9}},}\ }\href@noop {} {\bibfield  {journal} {\bibinfo
  {journal} {J. Chem. Theory Comput.}\ }\textbf {\bibinfo {volume} {9}},\
  \bibinfo {pages} {2000--2009} (\bibinfo {year} {2013})}\BibitemShut {NoStop}%
\bibitem [{\citenamefont {P{\'e}rez-Hern{\'a}ndez}\ \emph
  {et~al.}(2013)\citenamefont {P{\'e}rez-Hern{\'a}ndez}, \citenamefont {Paul},
  \citenamefont {Giorgino}, \citenamefont {De~Fabritiis},\ and\ \citenamefont
  {No{\'e}}}]{perez2013identification}%
  \BibitemOpen
  \bibfield  {author} {\bibinfo {author} {\bibfnamefont {G.}~\bibnamefont
  {P{\'e}rez-Hern{\'a}ndez}}, \bibinfo {author} {\bibfnamefont
  {F.}~\bibnamefont {Paul}}, \bibinfo {author} {\bibfnamefont {T.}~\bibnamefont
  {Giorgino}}, \bibinfo {author} {\bibfnamefont {G.}~\bibnamefont
  {De~Fabritiis}}, \ and\ \bibinfo {author} {\bibfnamefont {F.}~\bibnamefont
  {No{\'e}}},\ }\bibfield  {title} {\enquote {\bibinfo {title} {Identification
  of slow molecular order parameters for {Markov} model construction},}\
  }\href@noop {} {\bibfield  {journal} {\bibinfo  {journal} {J. Chem. Phys.}\
  }\textbf {\bibinfo {volume} {139}},\ \bibinfo {pages} {015102} (\bibinfo
  {year} {2013})}\BibitemShut {NoStop}%
\bibitem [{Note1()}]{Note1}%
  \BibitemOpen
  \bibinfo {note} {Here, we adopt the statistics convention by writing our
  linear equation $\protect \mathbf {Y} = \protect \mathbf {X}\protect \mathbf
  {K}$. In the propagator context, it is more common to write $\protect \mathbf
  {Y}=\protect \mathbf {K}\protect \mathbf {X}$, and thus the equations for
  $\protect \mathbf {K}$ in this work are equal to $\protect \mathbf {K}^\top $
  with the latter convention.}\BibitemShut {Stop}%
\bibitem [{\citenamefont {Paul}\ \emph {et~al.}(2019)\citenamefont {Paul},
  \citenamefont {Wu}, \citenamefont {Vossel}, \citenamefont {de~Groot},\ and\
  \citenamefont {No{\'e}}}]{paul2019identification}%
  \BibitemOpen
  \bibfield  {author} {\bibinfo {author} {\bibfnamefont {F.}~\bibnamefont
  {Paul}}, \bibinfo {author} {\bibfnamefont {H.}~\bibnamefont {Wu}}, \bibinfo
  {author} {\bibfnamefont {M.}~\bibnamefont {Vossel}}, \bibinfo {author}
  {\bibfnamefont {B.~L.}\ \bibnamefont {de~Groot}}, \ and\ \bibinfo {author}
  {\bibfnamefont {F.}~\bibnamefont {No{\'e}}},\ }\bibfield  {title} {\enquote
  {\bibinfo {title} {Identification of kinetic order parameters for
  non-equilibrium dynamics},}\ }\href@noop {} {\bibfield  {journal} {\bibinfo
  {journal} {J. Chem. Phys.}\ }\textbf {\bibinfo {volume} {150}},\ \bibinfo
  {pages} {164120} (\bibinfo {year} {2019})}\BibitemShut {NoStop}%
\bibitem [{\citenamefont {Wehmeyer}\ and\ \citenamefont
  {No{\'e}}(2018)}]{wehmeyer2018time}%
  \BibitemOpen
  \bibfield  {author} {\bibinfo {author} {\bibfnamefont {C.}~\bibnamefont
  {Wehmeyer}}\ and\ \bibinfo {author} {\bibfnamefont {F.}~\bibnamefont
  {No{\'e}}},\ }\bibfield  {title} {\enquote {\bibinfo {title} {Time-lagged
  autoencoders: Deep learning of slow collective variables for molecular
  kinetics},}\ }\href@noop {} {\bibfield  {journal} {\bibinfo  {journal} {J.
  Chem. Phys.}\ }\textbf {\bibinfo {volume} {148}},\ \bibinfo {pages} {241703}
  (\bibinfo {year} {2018})}\BibitemShut {NoStop}%
\bibitem [{Note2()}]{Note2}%
  \BibitemOpen
  \bibinfo {note} {We assume in~ \protect \textup {\hbox {\mathsurround \z@
  \protect \normalfont (\ignorespaces \ref {eq:xtilde}\unskip \@@italiccorr
  )}}-\protect \textup {\hbox {\mathsurround \z@ \protect \normalfont
  (\ignorespaces \ref {eq:ktilde}\unskip \@@italiccorr )}} that the matrices
  $\protect \mathbf {X}^\top \protect \mathbf {X}$, etc., are full
  rank.}\BibitemShut {Stop}%
\bibitem [{\citenamefont {Hotelling}(1936)}]{hotelling1936relations}%
  \BibitemOpen
  \bibfield  {author} {\bibinfo {author} {\bibfnamefont {H.}~\bibnamefont
  {Hotelling}},\ }\bibfield  {title} {\enquote {\bibinfo {title} {Relations
  between two sets of variates},}\ }\href@noop {} {\bibfield  {journal}
  {\bibinfo  {journal} {Biometrika}\ }\textbf {\bibinfo {volume} {28}},\
  \bibinfo {pages} {321--377} (\bibinfo {year} {1936})}\BibitemShut {NoStop}%
\bibitem [{\citenamefont {Wold}(2006)}]{wold1985partial}%
  \BibitemOpen
  \bibfield  {author} {\bibinfo {author} {\bibfnamefont {H.}~\bibnamefont
  {Wold}},\ }\enquote {\bibinfo {title} {Partial least squares},}\ in\ \href
  {\doibase 10.1002/0471667196.ess1914.pub2} {\emph {\bibinfo {booktitle}
  {Encyclopedia of Statistical Sciences}}}\ (\bibinfo  {publisher} {American
  Cancer Society},\ \bibinfo {year} {2006})\BibitemShut {NoStop}%
\bibitem [{\citenamefont {No{\'e}}(2018)}]{noe2018machine}%
  \BibitemOpen
  \bibfield  {author} {\bibinfo {author} {\bibfnamefont {F.}~\bibnamefont
  {No{\'e}}},\ }\bibfield  {title} {\enquote {\bibinfo {title} {Machine
  learning for molecular dynamics on long timescales},}\ }\href@noop {}
  {\bibfield  {journal} {\bibinfo  {journal} {arXiv preprint arXiv:1812.07669}\
  } (\bibinfo {year} {2018})}\BibitemShut {NoStop}%
\bibitem [{Note3()}]{Note3}%
  \BibitemOpen
  \bibinfo {note} {The Moore-Penrose pseudoinverse $\protect \mathbf {A}^+$ is
  defined as $(\protect \mathbf {A}^\top \protect \mathbf {A})^{-1}\protect
  \mathbf {A}^\top $ and is equivalent to $\protect \mathbf {A}^{-1}$ when
  $\protect \mathbf {A}$ is invertible~\cite
  {penrose1955generalized}.}\BibitemShut {Stop}%
\bibitem [{\citenamefont {Wold}(1973)}]{wold1973nonlinear}%
  \BibitemOpen
  \bibfield  {author} {\bibinfo {author} {\bibfnamefont {H.}~\bibnamefont
  {Wold}},\ }\bibfield  {title} {\enquote {\bibinfo {title} {Nonlinear
  iterative partial least squares (nipals) modelling: some current
  developments},}\ }in\ \href@noop {} {\emph {\bibinfo {booktitle}
  {Multivariate Analysis--III}}}\ (\bibinfo  {publisher} {Elsevier},\ \bibinfo
  {year} {1973})\ pp.\ \bibinfo {pages} {383--407}\BibitemShut {NoStop}%
\bibitem [{\citenamefont {Wegelin}\ \emph {et~al.}(2000)\citenamefont {Wegelin}
  \emph {et~al.}}]{wegelin2000survey}%
  \BibitemOpen
  \bibfield  {author} {\bibinfo {author} {\bibfnamefont {J.~A.}\ \bibnamefont
  {Wegelin}} \emph {et~al.},\ }\bibfield  {title} {\enquote {\bibinfo {title}
  {A survey of partial least squares (pls) methods, with emphasis on the
  two-block case},}\ }\href@noop {} {\bibfield  {journal} {\bibinfo  {journal}
  {University of Washington, Department of Statistics, Tech. Rep}\ } (\bibinfo
  {year} {2000})}\BibitemShut {NoStop}%
\bibitem [{\citenamefont {Wold}(1975)}]{wold1975path}%
  \BibitemOpen
  \bibfield  {author} {\bibinfo {author} {\bibfnamefont {H.}~\bibnamefont
  {Wold}},\ }\bibfield  {title} {\enquote {\bibinfo {title} {Path models with
  latent variables: The {NIPALS} approach},}\ }in\ \href@noop {} {\emph
  {\bibinfo {booktitle} {Quantitative Sociology}}}\ (\bibinfo  {publisher}
  {Elsevier},\ \bibinfo {year} {1975})\ pp.\ \bibinfo {pages}
  {307--357}\BibitemShut {NoStop}%
\bibitem [{Note4()}]{Note4}%
  \BibitemOpen
  \bibinfo {note} {PCA is equivalent to PLS when $\protect \mathbf {X} =
  \protect \mathbf {Y}$.}\BibitemShut {Stop}%
\bibitem [{\citenamefont {White}(1958)}]{white1958computation}%
  \BibitemOpen
  \bibfield  {author} {\bibinfo {author} {\bibfnamefont {P.~A.}\ \bibnamefont
  {White}},\ }\bibfield  {title} {\enquote {\bibinfo {title} {The computation
  of eigenvalues and eigenvectors of a matrix},}\ }\href@noop {} {\bibfield
  {journal} {\bibinfo  {journal} {SIAM J. Appl. Math.}\ }\textbf {\bibinfo
  {volume} {6}},\ \bibinfo {pages} {393--437} (\bibinfo {year}
  {1958})}\BibitemShut {NoStop}%
\bibitem [{Note5()}]{Note5}%
  \BibitemOpen
  \bibinfo {note} {Note that if the deflated weights are applied to the the
  deflated data, the transformations will be the same in VAMP and dVAMP. The
  nonequivalence of VAMP and dVAMP transformations occurs when applying the
  deflated weights to \protect \emph {nondeflated} data.}\BibitemShut {Stop}%
\bibitem [{\citenamefont {Spearman}(1904)}]{spearman1904proof}%
  \BibitemOpen
  \bibfield  {author} {\bibinfo {author} {\bibfnamefont {C.}~\bibnamefont
  {Spearman}},\ }\bibfield  {title} {\enquote {\bibinfo {title} {The proof and
  measurement of association between two things},}\ }\href
  {http://www.jstor.org/stable/1412159} {\bibfield  {journal} {\bibinfo
  {journal} {Am. J. Psychol.}\ }\textbf {\bibinfo {volume} {15}},\ \bibinfo
  {pages} {72--101} (\bibinfo {year} {1904})}\BibitemShut {NoStop}%
\bibitem [{\citenamefont {Lindorff-Larsen}\ \emph {et~al.}(2011)\citenamefont
  {Lindorff-Larsen}, \citenamefont {Piana}, \citenamefont {Dror},\ and\
  \citenamefont {Shaw}}]{lindorff2011fast}%
  \BibitemOpen
  \bibfield  {author} {\bibinfo {author} {\bibfnamefont {K.}~\bibnamefont
  {Lindorff-Larsen}}, \bibinfo {author} {\bibfnamefont {S.}~\bibnamefont
  {Piana}}, \bibinfo {author} {\bibfnamefont {R.~O.}\ \bibnamefont {Dror}}, \
  and\ \bibinfo {author} {\bibfnamefont {D.~E.}\ \bibnamefont {Shaw}},\
  }\bibfield  {title} {\enquote {\bibinfo {title} {How fast-folding proteins
  fold},}\ }\href@noop {} {\bibfield  {journal} {\bibinfo  {journal} {Science}\
  }\textbf {\bibinfo {volume} {334}},\ \bibinfo {pages} {517--520} (\bibinfo
  {year} {2011})}\BibitemShut {NoStop}%
\bibitem [{\citenamefont {No{\'e}}\ and\ \citenamefont
  {Clementi}(2015)}]{noe2015kinetic}%
  \BibitemOpen
  \bibfield  {author} {\bibinfo {author} {\bibfnamefont {F.}~\bibnamefont
  {No{\'e}}}\ and\ \bibinfo {author} {\bibfnamefont {C.}~\bibnamefont
  {Clementi}},\ }\bibfield  {title} {\enquote {\bibinfo {title} {Kinetic
  distance and kinetic maps from molecular dynamics simulation},}\ }\href@noop
  {} {\bibfield  {journal} {\bibinfo  {journal} {J. Chem. Theory Comput.}\
  }\textbf {\bibinfo {volume} {11}},\ \bibinfo {pages} {5002--5011} (\bibinfo
  {year} {2015})}\BibitemShut {NoStop}%
\bibitem [{\citenamefont {McGibbon}\ and\ \citenamefont
  {Pande}(2015)}]{mcgibbon2015variational}%
  \BibitemOpen
  \bibfield  {author} {\bibinfo {author} {\bibfnamefont {R.~T.}\ \bibnamefont
  {McGibbon}}\ and\ \bibinfo {author} {\bibfnamefont {V.~S.}\ \bibnamefont
  {Pande}},\ }\bibfield  {title} {\enquote {\bibinfo {title} {Variational
  cross-validation of slow dynamical modes in molecular kinetics},}\
  }\href@noop {} {\bibfield  {journal} {\bibinfo  {journal} {J. Chem. Phys.}\
  }\textbf {\bibinfo {volume} {142}},\ \bibinfo {pages} {124105} (\bibinfo
  {year} {2015})}\BibitemShut {NoStop}%
\bibitem [{\citenamefont {Deuflhard}\ \emph {et~al.}(2000)\citenamefont
  {Deuflhard}, \citenamefont {Huisinga}, \citenamefont {Fischer},\ and\
  \citenamefont {Sch{\"u}tte}}]{deuflhard2000identification}%
  \BibitemOpen
  \bibfield  {author} {\bibinfo {author} {\bibfnamefont {P.}~\bibnamefont
  {Deuflhard}}, \bibinfo {author} {\bibfnamefont {W.}~\bibnamefont {Huisinga}},
  \bibinfo {author} {\bibfnamefont {A.}~\bibnamefont {Fischer}}, \ and\
  \bibinfo {author} {\bibfnamefont {C.}~\bibnamefont {Sch{\"u}tte}},\
  }\bibfield  {title} {\enquote {\bibinfo {title} {Identification of almost
  invariant aggregates in reversible nearly uncoupled markov chains},}\
  }\href@noop {} {\bibfield  {journal} {\bibinfo  {journal} {Linear Alg.
  Appl.}\ }\textbf {\bibinfo {volume} {315}},\ \bibinfo {pages} {39--59}
  (\bibinfo {year} {2000})}\BibitemShut {NoStop}%
\bibitem [{Note6()}]{Note6}%
  \BibitemOpen
  \bibinfo {note} {This is different from $\protect \mathbf {\xi }'_c$ in
  Eqn.~\ref {eq:scorex}, which is not calculated in the course of Alg.~\ref
  {alg:ccanipals} on data that has been deflated $c$ times, but instead is
  calculated on the original data afterward.}\BibitemShut {Stop}%
\bibitem [{Note7()}]{Note7}%
  \BibitemOpen
  \bibinfo {note} {Note that if the undesired process is not the slowest
  process, i.e., $c > 1$, we can deflate it exclusively using~\protect \textup
  {\hbox {\mathsurround \z@ \protect \normalfont (\ignorespaces \ref
  {eq:deflation}\unskip \@@italiccorr )}} once for $c$ only. However, we must
  perform dVAMP \protect \emph {through} the $c$th coordinates to obtain the
  necessary weights.}\BibitemShut {Stop}%
\bibitem [{Note8()}]{Note8}%
  \BibitemOpen
  \bibinfo {note} {To calculate all of the distances, we perform the same
  transformation for $\protect \mathbf {Y}$ using $\omega _c$ in order to
  obtain the last $\tau $ data points, which are not present in $\protect
  \mathbf {X}$ due to the time lag.}\BibitemShut {Stop}%
\bibitem [{\citenamefont {Beauchamp}\ \emph {et~al.}(2012)\citenamefont
  {Beauchamp}, \citenamefont {McGibbon}, \citenamefont {Lin},\ and\
  \citenamefont {Pande}}]{beauchamp2012simple}%
  \BibitemOpen
  \bibfield  {author} {\bibinfo {author} {\bibfnamefont {K.~A.}\ \bibnamefont
  {Beauchamp}}, \bibinfo {author} {\bibfnamefont {R.}~\bibnamefont {McGibbon}},
  \bibinfo {author} {\bibfnamefont {Y.-S.}\ \bibnamefont {Lin}}, \ and\
  \bibinfo {author} {\bibfnamefont {V.~S.}\ \bibnamefont {Pande}},\ }\bibfield
  {title} {\enquote {\bibinfo {title} {Simple few-state models reveal hidden
  complexity in protein folding},}\ }\href@noop {} {\bibfield  {journal}
  {\bibinfo  {journal} {Proc. Natl. Acad. Sci.}\ }\textbf {\bibinfo {volume}
  {109}},\ \bibinfo {pages} {17807--17813} (\bibinfo {year}
  {2012})}\BibitemShut {NoStop}%
\bibitem [{\citenamefont {Hub}\ and\ \citenamefont
  {De~Groot}(2009)}]{hub2009detection}%
  \BibitemOpen
  \bibfield  {author} {\bibinfo {author} {\bibfnamefont {J.~S.}\ \bibnamefont
  {Hub}}\ and\ \bibinfo {author} {\bibfnamefont {B.~L.}\ \bibnamefont
  {De~Groot}},\ }\bibfield  {title} {\enquote {\bibinfo {title} {Detection of
  functional modes in protein dynamics},}\ }\href@noop {} {\bibfield  {journal}
  {\bibinfo  {journal} {PLoS Comput. Biol.}\ }\textbf {\bibinfo {volume} {5}},\
  \bibinfo {pages} {e1000480} (\bibinfo {year} {2009})}\BibitemShut {NoStop}%
\bibitem [{\citenamefont {Krivobokova}\ \emph {et~al.}(2012)\citenamefont
  {Krivobokova}, \citenamefont {Briones}, \citenamefont {Hub}, \citenamefont
  {Munk},\ and\ \citenamefont {de~Groot}}]{krivobokova2012partial}%
  \BibitemOpen
  \bibfield  {author} {\bibinfo {author} {\bibfnamefont {T.}~\bibnamefont
  {Krivobokova}}, \bibinfo {author} {\bibfnamefont {R.}~\bibnamefont
  {Briones}}, \bibinfo {author} {\bibfnamefont {J.~S.}\ \bibnamefont {Hub}},
  \bibinfo {author} {\bibfnamefont {A.}~\bibnamefont {Munk}}, \ and\ \bibinfo
  {author} {\bibfnamefont {B.~L.}\ \bibnamefont {de~Groot}},\ }\bibfield
  {title} {\enquote {\bibinfo {title} {Partial least-squares functional mode
  analysis: application to the membrane proteins aqp1, aqy1, and clc-ec1},}\
  }\href@noop {} {\bibfield  {journal} {\bibinfo  {journal} {Biophys. J.}\
  }\textbf {\bibinfo {volume} {103}},\ \bibinfo {pages} {786--796} (\bibinfo
  {year} {2012})}\BibitemShut {NoStop}%
\bibitem [{\citenamefont {Wehmeyer}\ \emph {et~al.}(2018)\citenamefont
  {Wehmeyer}, \citenamefont {Scherer}, \citenamefont {Hempel}, \citenamefont
  {Husic}, \citenamefont {Olsson},\ and\ \citenamefont
  {No{\'e}}}]{wehmeyer2018introduction}%
  \BibitemOpen
  \bibfield  {author} {\bibinfo {author} {\bibfnamefont {C.}~\bibnamefont
  {Wehmeyer}}, \bibinfo {author} {\bibfnamefont {M.~K.}\ \bibnamefont
  {Scherer}}, \bibinfo {author} {\bibfnamefont {T.}~\bibnamefont {Hempel}},
  \bibinfo {author} {\bibfnamefont {B.~E.}\ \bibnamefont {Husic}}, \bibinfo
  {author} {\bibfnamefont {S.}~\bibnamefont {Olsson}}, \ and\ \bibinfo {author}
  {\bibfnamefont {F.}~\bibnamefont {No{\'e}}},\ }\bibfield  {title} {\enquote
  {\bibinfo {title} {Introduction to {Markov} state modeling with the {PyEMMA}
  software [article v1.0]},}\ }\href {\doibase 10.33011/livecoms.1.1.5965}
  {\bibfield  {journal} {\bibinfo  {journal} {LiveCoMS}\ }\textbf {\bibinfo
  {volume} {1}},\ \bibinfo {pages} {5965} (\bibinfo {year} {2018})}\BibitemShut
  {NoStop}%
\bibitem [{\citenamefont {Foreman-Mackey}(2016)}]{foreman2016corner}%
  \BibitemOpen
  \bibfield  {author} {\bibinfo {author} {\bibfnamefont {D.}~\bibnamefont
  {Foreman-Mackey}},\ }\bibfield  {title} {\enquote {\bibinfo {title}
  {corner.py: Scatterplot matrices in python},}\ }\href
  {http://dx.doi.org/10.5281/zenodo.45906} {\bibfield  {journal} {\bibinfo
  {journal} {J. Open Source Softw.}\ }\textbf {\bibinfo {volume} {24}}
  (\bibinfo {year} {2016})}\BibitemShut {NoStop}%
\bibitem [{\citenamefont {McGibbon}\ \emph {et~al.}(2015)\citenamefont
  {McGibbon}, \citenamefont {Beauchamp}, \citenamefont {Harrigan},
  \citenamefont {Klein}, \citenamefont {Swails}, \citenamefont {Hern{\'a}ndez},
  \citenamefont {Schwantes}, \citenamefont {Wang}, \citenamefont {Lane},\ and\
  \citenamefont {Pande}}]{mcgibbon2015mdtraj}%
  \BibitemOpen
  \bibfield  {author} {\bibinfo {author} {\bibfnamefont {R.~T.}\ \bibnamefont
  {McGibbon}}, \bibinfo {author} {\bibfnamefont {K.~A.}\ \bibnamefont
  {Beauchamp}}, \bibinfo {author} {\bibfnamefont {M.~P.}\ \bibnamefont
  {Harrigan}}, \bibinfo {author} {\bibfnamefont {C.}~\bibnamefont {Klein}},
  \bibinfo {author} {\bibfnamefont {J.~M.}\ \bibnamefont {Swails}}, \bibinfo
  {author} {\bibfnamefont {C.~X.}\ \bibnamefont {Hern{\'a}ndez}}, \bibinfo
  {author} {\bibfnamefont {C.~R.}\ \bibnamefont {Schwantes}}, \bibinfo {author}
  {\bibfnamefont {L.-P.}\ \bibnamefont {Wang}}, \bibinfo {author}
  {\bibfnamefont {T.~J.}\ \bibnamefont {Lane}}, \ and\ \bibinfo {author}
  {\bibfnamefont {V.~S.}\ \bibnamefont {Pande}},\ }\bibfield  {title} {\enquote
  {\bibinfo {title} {{MDTraj}: a modern open library for the analysis of
  molecular dynamics trajectories},}\ }\href@noop {} {\bibfield  {journal}
  {\bibinfo  {journal} {Biophys. J.}\ }\textbf {\bibinfo {volume} {109}},\
  \bibinfo {pages} {1528--1532} (\bibinfo {year} {2015})}\BibitemShut {NoStop}%
\bibitem [{\citenamefont {Kluyver}\ \emph {et~al.}(2016)\citenamefont
  {Kluyver}, \citenamefont {Ragan-Kelley}, \citenamefont {P{\'e}rez},
  \citenamefont {Granger}, \citenamefont {Bussonnier}, \citenamefont
  {Frederic}, \citenamefont {Kelley}, \citenamefont {Hamrick}, \citenamefont
  {Grout}, \citenamefont {Corlay}, \citenamefont {Ivanov}, \citenamefont
  {Avila}, \citenamefont {Abdalla},\ and\ \citenamefont {Willing}}]{jupyter}%
  \BibitemOpen
  \bibfield  {author} {\bibinfo {author} {\bibfnamefont {T.}~\bibnamefont
  {Kluyver}}, \bibinfo {author} {\bibfnamefont {B.}~\bibnamefont
  {Ragan-Kelley}}, \bibinfo {author} {\bibfnamefont {F.}~\bibnamefont
  {P{\'e}rez}}, \bibinfo {author} {\bibfnamefont {B.}~\bibnamefont {Granger}},
  \bibinfo {author} {\bibfnamefont {M.}~\bibnamefont {Bussonnier}}, \bibinfo
  {author} {\bibfnamefont {J.}~\bibnamefont {Frederic}}, \bibinfo {author}
  {\bibfnamefont {K.}~\bibnamefont {Kelley}}, \bibinfo {author} {\bibfnamefont
  {J.}~\bibnamefont {Hamrick}}, \bibinfo {author} {\bibfnamefont
  {J.}~\bibnamefont {Grout}}, \bibinfo {author} {\bibfnamefont
  {S.}~\bibnamefont {Corlay}}, \bibinfo {author} {\bibfnamefont
  {P.}~\bibnamefont {Ivanov}}, \bibinfo {author} {\bibfnamefont
  {D.}~\bibnamefont {Avila}}, \bibinfo {author} {\bibfnamefont
  {S.}~\bibnamefont {Abdalla}}, \ and\ \bibinfo {author} {\bibfnamefont
  {C.}~\bibnamefont {Willing}},\ }\bibfield  {title} {\enquote {\bibinfo
  {title} {Jupyter notebooks -- a publishing format for reproducible
  computational workflows},}\ }in\ \href@noop {} {\emph {\bibinfo {booktitle}
  {Positioning and Power in Academic Publishing: Players, Agents and
  Agendas}}},\ \bibinfo {editor} {edited by\ \bibinfo {editor} {\bibfnamefont
  {F.}~\bibnamefont {Loizides}}\ and\ \bibinfo {editor} {\bibfnamefont
  {B.}~\bibnamefont {Schmidt}}}\ (\bibinfo {organization} {IOS Press},\
  \bibinfo {year} {2016})\ pp.\ \bibinfo {pages} {87--90}\BibitemShut {NoStop}%
\bibitem [{\citenamefont {Van Der~Walt}, \citenamefont {Colbert},\ and\
  \citenamefont {Varoquaux}(2011)}]{numpy}%
  \BibitemOpen
  \bibfield  {author} {\bibinfo {author} {\bibfnamefont {S.}~\bibnamefont {Van
  Der~Walt}}, \bibinfo {author} {\bibfnamefont {S.~C.}\ \bibnamefont
  {Colbert}}, \ and\ \bibinfo {author} {\bibfnamefont {G.}~\bibnamefont
  {Varoquaux}},\ }\bibfield  {title} {\enquote {\bibinfo {title} {The numpy
  array: a structure for efficient numerical computation},}\ }\href@noop {}
  {\bibfield  {journal} {\bibinfo  {journal} {Comput. Sci. Eng.}\ }\textbf
  {\bibinfo {volume} {13}},\ \bibinfo {pages} {22} (\bibinfo {year}
  {2011})}\BibitemShut {NoStop}%
\bibitem [{\citenamefont {Jones}\ \emph {et~al.}(01  )\citenamefont {Jones},
  \citenamefont {Oliphant}, \citenamefont {Peterson} \emph {et~al.}}]{scipy}%
  \BibitemOpen
  \bibfield  {author} {\bibinfo {author} {\bibfnamefont {E.}~\bibnamefont
  {Jones}}, \bibinfo {author} {\bibfnamefont {T.}~\bibnamefont {Oliphant}},
  \bibinfo {author} {\bibfnamefont {P.}~\bibnamefont {Peterson}},  \emph
  {et~al.},\ }\href {http://www.scipy.org/} {\enquote {\bibinfo {title}
  {{SciPy}: Open source scientific tools for {Python}},}\ } (\bibinfo {year}
  {2001--})\BibitemShut {NoStop}%
\bibitem [{\citenamefont {Pedregosa}\ \emph {et~al.}(2011)\citenamefont
  {Pedregosa}, \citenamefont {Varoquaux}, \citenamefont {Gramfort},
  \citenamefont {Michel}, \citenamefont {Thirion}, \citenamefont {Grisel},
  \citenamefont {Blondel}, \citenamefont {Prettenhofer}, \citenamefont {Weiss},
  \citenamefont {Dubourg}, \citenamefont {Vanderplas}, \citenamefont {Passos},
  \citenamefont {Cournapeau}, \citenamefont {Brucher}, \citenamefont {Perrot},\
  and\ \citenamefont {Duchesnay}}]{scikit-learn}%
  \BibitemOpen
  \bibfield  {author} {\bibinfo {author} {\bibfnamefont {F.}~\bibnamefont
  {Pedregosa}}, \bibinfo {author} {\bibfnamefont {G.}~\bibnamefont
  {Varoquaux}}, \bibinfo {author} {\bibfnamefont {A.}~\bibnamefont {Gramfort}},
  \bibinfo {author} {\bibfnamefont {V.}~\bibnamefont {Michel}}, \bibinfo
  {author} {\bibfnamefont {B.}~\bibnamefont {Thirion}}, \bibinfo {author}
  {\bibfnamefont {O.}~\bibnamefont {Grisel}}, \bibinfo {author} {\bibfnamefont
  {M.}~\bibnamefont {Blondel}}, \bibinfo {author} {\bibfnamefont
  {P.}~\bibnamefont {Prettenhofer}}, \bibinfo {author} {\bibfnamefont
  {R.}~\bibnamefont {Weiss}}, \bibinfo {author} {\bibfnamefont
  {V.}~\bibnamefont {Dubourg}}, \bibinfo {author} {\bibfnamefont
  {J.}~\bibnamefont {Vanderplas}}, \bibinfo {author} {\bibfnamefont
  {A.}~\bibnamefont {Passos}}, \bibinfo {author} {\bibfnamefont
  {D.}~\bibnamefont {Cournapeau}}, \bibinfo {author} {\bibfnamefont
  {M.}~\bibnamefont {Brucher}}, \bibinfo {author} {\bibfnamefont
  {M.}~\bibnamefont {Perrot}}, \ and\ \bibinfo {author} {\bibfnamefont
  {E.}~\bibnamefont {Duchesnay}},\ }\bibfield  {title} {\enquote {\bibinfo
  {title} {Scikit-learn: Machine learning in {P}ython},}\ }\href@noop {}
  {\bibfield  {journal} {\bibinfo  {journal} {J. Mach. Learn. Res.}\ }\textbf
  {\bibinfo {volume} {12}},\ \bibinfo {pages} {2825--2830} (\bibinfo {year}
  {2011})}\BibitemShut {NoStop}%
\bibitem [{\citenamefont {Hunter}(2007)}]{matplotlib}%
  \BibitemOpen
  \bibfield  {author} {\bibinfo {author} {\bibfnamefont {J.~D.}\ \bibnamefont
  {Hunter}},\ }\bibfield  {title} {\enquote {\bibinfo {title} {Matplotlib: A
  {2D} graphics environment},}\ }\href {\doibase 10.1109/MCSE.2007.55}
  {\bibfield  {journal} {\bibinfo  {journal} {Comput. Sci. Eng.}\ }\textbf
  {\bibinfo {volume} {9}},\ \bibinfo {pages} {90--95} (\bibinfo {year}
  {2007})}\BibitemShut {NoStop}%
\bibitem [{\citenamefont {Waskom}\ \emph {et~al.}(2017)\citenamefont {Waskom},
  \citenamefont {Botvinnik}, \citenamefont {O'Kane}, \citenamefont {Hobson},
  \citenamefont {Lukauskas}, \citenamefont {Gemperline}, \citenamefont
  {Augspurger}, \citenamefont {Halchenko}, \citenamefont {Cole}, \citenamefont
  {Warmenhoven}, \citenamefont {de~Ruiter}, \citenamefont {Pye}, \citenamefont
  {Hoyer}, \citenamefont {Vanderplas}, \citenamefont {Villalba}, \citenamefont
  {Kunter}, \citenamefont {Quintero}, \citenamefont {Bachant}, \citenamefont
  {Martin}, \citenamefont {Meyer}, \citenamefont {Miles}, \citenamefont {Ram},
  \citenamefont {Yarkoni}, \citenamefont {Williams}, \citenamefont {Evans},
  \citenamefont {Fitzgerald}, \citenamefont {Brian}, \citenamefont
  {Fonnesbeck}, \citenamefont {Lee},\ and\ \citenamefont {Qalieh}}]{seaborn}%
  \BibitemOpen
  \bibfield  {author} {\bibinfo {author} {\bibfnamefont {M.}~\bibnamefont
  {Waskom}}, \bibinfo {author} {\bibfnamefont {O.}~\bibnamefont {Botvinnik}},
  \bibinfo {author} {\bibfnamefont {D.}~\bibnamefont {O'Kane}}, \bibinfo
  {author} {\bibfnamefont {P.}~\bibnamefont {Hobson}}, \bibinfo {author}
  {\bibfnamefont {S.}~\bibnamefont {Lukauskas}}, \bibinfo {author}
  {\bibfnamefont {D.~C.}\ \bibnamefont {Gemperline}}, \bibinfo {author}
  {\bibfnamefont {T.}~\bibnamefont {Augspurger}}, \bibinfo {author}
  {\bibfnamefont {Y.}~\bibnamefont {Halchenko}}, \bibinfo {author}
  {\bibfnamefont {J.~B.}\ \bibnamefont {Cole}}, \bibinfo {author}
  {\bibfnamefont {J.}~\bibnamefont {Warmenhoven}}, \bibinfo {author}
  {\bibfnamefont {J.}~\bibnamefont {de~Ruiter}}, \bibinfo {author}
  {\bibfnamefont {C.}~\bibnamefont {Pye}}, \bibinfo {author} {\bibfnamefont
  {S.}~\bibnamefont {Hoyer}}, \bibinfo {author} {\bibfnamefont
  {J.}~\bibnamefont {Vanderplas}}, \bibinfo {author} {\bibfnamefont
  {S.}~\bibnamefont {Villalba}}, \bibinfo {author} {\bibfnamefont
  {G.}~\bibnamefont {Kunter}}, \bibinfo {author} {\bibfnamefont
  {E.}~\bibnamefont {Quintero}}, \bibinfo {author} {\bibfnamefont
  {P.}~\bibnamefont {Bachant}}, \bibinfo {author} {\bibfnamefont
  {M.}~\bibnamefont {Martin}}, \bibinfo {author} {\bibfnamefont
  {K.}~\bibnamefont {Meyer}}, \bibinfo {author} {\bibfnamefont
  {A.}~\bibnamefont {Miles}}, \bibinfo {author} {\bibfnamefont
  {Y.}~\bibnamefont {Ram}}, \bibinfo {author} {\bibfnamefont {T.}~\bibnamefont
  {Yarkoni}}, \bibinfo {author} {\bibfnamefont {M.~L.}\ \bibnamefont
  {Williams}}, \bibinfo {author} {\bibfnamefont {C.}~\bibnamefont {Evans}},
  \bibinfo {author} {\bibfnamefont {C.}~\bibnamefont {Fitzgerald}}, \bibinfo
  {author} {\bibnamefont {Brian}}, \bibinfo {author} {\bibfnamefont
  {C.}~\bibnamefont {Fonnesbeck}}, \bibinfo {author} {\bibfnamefont
  {A.}~\bibnamefont {Lee}}, \ and\ \bibinfo {author} {\bibfnamefont
  {A.}~\bibnamefont {Qalieh}},\ }\href {\doibase 10.5281/zenodo.883859}
  {\enquote {\bibinfo {title} {mwaskom/seaborn: v0.8.1 (september 2017)},}\ }
  (\bibinfo {year} {2017})\BibitemShut {NoStop}%
\bibitem [{\citenamefont {Humphrey}, \citenamefont {Dalke},\ and\ \citenamefont
  {Schulten}(1996)}]{vmd}%
  \BibitemOpen
  \bibfield  {author} {\bibinfo {author} {\bibfnamefont {W.}~\bibnamefont
  {Humphrey}}, \bibinfo {author} {\bibfnamefont {A.}~\bibnamefont {Dalke}}, \
  and\ \bibinfo {author} {\bibfnamefont {K.}~\bibnamefont {Schulten}},\
  }\bibfield  {title} {\enquote {\bibinfo {title} {{VMD} -- {V}isual
  {M}olecular {D}ynamics},}\ }\href@noop {} {\bibfield  {journal} {\bibinfo
  {journal} {J. Mol. Graph.}\ }\textbf {\bibinfo {volume} {14}},\ \bibinfo
  {pages} {33--38} (\bibinfo {year} {1996})}\BibitemShut {NoStop}%
\bibitem [{\citenamefont {Penrose}(1955)}]{penrose1955generalized}%
  \BibitemOpen
  \bibfield  {author} {\bibinfo {author} {\bibfnamefont {R.}~\bibnamefont
  {Penrose}},\ }\bibfield  {title} {\enquote {\bibinfo {title} {A generalized
  inverse for matrices},}\ }\href {\doibase 10.1017/S0305004100030401}
  {\bibfield  {journal} {\bibinfo  {journal} {Math. Proc. Cambridge Philos.
  Soc.}\ }\textbf {\bibinfo {volume} {51}},\ \bibinfo {pages} {406–413}
  (\bibinfo {year} {1955})}\BibitemShut {NoStop}%
\bibitem [{\citenamefont {Koopman}(1931)}]{koopman1931hamiltonian}%
  \BibitemOpen
  \bibfield  {author} {\bibinfo {author} {\bibfnamefont {B.~O.}\ \bibnamefont
  {Koopman}},\ }\bibfield  {title} {\enquote {\bibinfo {title} {Hamiltonian
  systems and transformation in hilbert space},}\ }\href@noop {} {\bibfield
  {journal} {\bibinfo  {journal} {Proc. Natl. Acad. Sci.}\ }\textbf {\bibinfo
  {volume} {17}},\ \bibinfo {pages} {315--318} (\bibinfo {year}
  {1931})}\BibitemShut {NoStop}%
\bibitem [{\citenamefont {Mezi{\'c}}(2005)}]{mezic2005spectral}%
  \BibitemOpen
  \bibfield  {author} {\bibinfo {author} {\bibfnamefont {I.}~\bibnamefont
  {Mezi{\'c}}},\ }\bibfield  {title} {\enquote {\bibinfo {title} {Spectral
  properties of dynamical systems, model reduction and decompositions},}\
  }\href@noop {} {\bibfield  {journal} {\bibinfo  {journal} {Nonlinear Dyn.}\
  }\textbf {\bibinfo {volume} {41}},\ \bibinfo {pages} {309--325} (\bibinfo
  {year} {2005})}\BibitemShut {NoStop}%
\bibitem [{\citenamefont {Klus}\ \emph {et~al.}(2018)\citenamefont {Klus},
  \citenamefont {N{\"u}ske}, \citenamefont {Koltai}, \citenamefont {Wu},
  \citenamefont {Kevrekidis}, \citenamefont {Sch{\"u}tte},\ and\ \citenamefont
  {No{\'e}}}]{klus2018data}%
  \BibitemOpen
  \bibfield  {author} {\bibinfo {author} {\bibfnamefont {S.}~\bibnamefont
  {Klus}}, \bibinfo {author} {\bibfnamefont {F.}~\bibnamefont {N{\"u}ske}},
  \bibinfo {author} {\bibfnamefont {P.}~\bibnamefont {Koltai}}, \bibinfo
  {author} {\bibfnamefont {H.}~\bibnamefont {Wu}}, \bibinfo {author}
  {\bibfnamefont {I.}~\bibnamefont {Kevrekidis}}, \bibinfo {author}
  {\bibfnamefont {C.}~\bibnamefont {Sch{\"u}tte}}, \ and\ \bibinfo {author}
  {\bibfnamefont {F.}~\bibnamefont {No{\'e}}},\ }\bibfield  {title} {\enquote
  {\bibinfo {title} {Data-driven model reduction and transfer operator
  approximation},}\ }\href@noop {} {\bibfield  {journal} {\bibinfo  {journal}
  {J. Nonlinear Sci.}\ }\textbf {\bibinfo {volume} {28}},\ \bibinfo {pages}
  {985--1010} (\bibinfo {year} {2018})}\BibitemShut {NoStop}%
\end{thebibliography}%

\end{document}